\documentclass[11pt,a4paper]{article}
\pdfoutput=1

\usepackage[utf8]{inputenc}
\usepackage{jheppub}
\usepackage{hyperref}

\usepackage{tikz} 
\usetikzlibrary{arrows} 
\usetikzlibrary{shapes,arrows,positioning,automata,backgrounds,calc,er,patterns}
\usepackage{tikz-feynman}
\tikzfeynmanset{compat=1.1.0}

\preprint{ \vbox{\hbox{IPARCOS-UCM-23-123}}}

\usepackage{color}
\usepackage{subfigure}
\usepackage{yhmath}

\definecolor{mygreen}{RGB}{0,128,0} 
\definecolor{mybrown}{RGB}{153,102,51}

\def\gsim{~\,\makebox(1,1){$\stackrel{>}{\widetilde{}}$}\,~}
\def\lsim{~\,\makebox(1,1){$\stackrel{<}{\widetilde{}}$}\,~}

\newcommand{\mF}{\mathcal{F}}
\newcommand{\mL}{\mathcal{L}}
\newcommand{\mM}{\mathcal{M}}
\newcommand{\mN}{\mathcal{N}}
\newcommand{\mO}{\mathcal{O}}
\newcommand{\mV}{\mathcal{V}}

\title{   Production of two, three, and four Higgs bosons: where SMEFT and HEFT depart }

\author[a]{Rafael L. Delgado,}
\author[b]{Raquel G\'omez-Ambrosio,}
\author[c]{Javier Mart\'\i nez-Mart\'\i n,}
\author[c,d]{Alexandre Salas-Bern\'ardez,}
\author[c]{Juan J. Sanz-Cillero}

\affiliation[a]{Dept. Matem\'atica Aplicadas a las TIC, Universidad Polit\'ecnica de Madrid, Nikola Tesla, s/n, 28031-Madrid, Spain}
\affiliation[b]{Dipartimento di Fisica, Universit\`a di Torino, and INFN, Sezione di Torino,
Via P. Giuria 1, 10125 Torino, Italy}
\affiliation[c]{Dept. F\'\i sica Te\'orica and IPARCOS, Universidad Complutense de Madrid, Plaza de las Ciencias 1, 28040 Madrid, Spain}
\affiliation[d]{Instituto de Física Corpuscular (IFIC), Universidad de Valencia-CSIC,
E-46980 Valencia, Spain}

\emailAdd{rafael.delgado@upm.es}
\emailAdd{raquel.gomezambrosio@unito.it}
\emailAdd{javmar21@ucm.es}
\emailAdd{alexsala@ucm.es}
\emailAdd{asaber2@uv.es}
\emailAdd{jjsanzcillero@ucm.es}

\abstract{ 
In this article we study the phenomenological implications of multiple Higgs boson production from longitudinal vector boson scattering in the context of effective field theories. 
We find compact representations for effective tree-level amplitudes with up to four final state Higgs bosons. Total cross sections are then computed for scenarios relevant at the LHC in which we find the general Higgs Effective Theory (HEFT) prediction avoids the heavy suppression observed in Standard Model Effective Field Theory (SMEFT). 
}

\date{\today}
\makeatletter
\gdef\@fpheader{}
\makeatother

\begin{document}
\maketitle 

\section{Introduction}

The Higgs boson was first observed in 2012 at the Large Hadron Collider (LHC)\cite{ATLAS:2012znl, CMS:2012zwa, CMS:2012qbp}, confirming the existence of a field that endows some of the other elementary particles with mass, playing a fundamental role in our understanding of the universe. However, many questions remain open regarding its origin and its interplay with the rest of the Standard Model (SM) fields. 

In recent years, effective field theories (EFTs)  have proven to be a useful tool to parametrize the low energy effects of the underlying, and yet unknown, ultraviolet physics. Among these EFTs, the Standard Model Effective Field Theory (SMEFT)  and Higgs Effective Field Theory (HEFT)  have emerged as powerful frameworks for describing particle interactions at various energy scales. In this paper, we explore the theoretical foundations and predictive capabilities of SMEFT and HEFT, with a particular focus on their applicability to the observed phenomena at the LHC. By comparing the predictions of these theories with experimental data obtained from the LHC, we aim to gain a deeper understanding of their strengths and limitations in describing the behaviour of subatomic particles. 

When trying to compare EFT predictions to experimental observations, two approaches
are the standard: either through a global fit of cross sections and differential distributions
or through the definition of realistic (pseudo-)observables, such as fiducial cross sections,
ratios, and asymmetries. Several groups have applied these techniques to the available data
from LHC Runs I and II, for example, see [4–15]. In the spirit of the latter approach, we propose a systematic study of longitudinal vector boson scattering cross sections.

Here, we focus on the electroweak sector of the Standard Model, allowing us to explore further the connection between the Higgs and gauge bosons.  In this sector, by only looking at a handful of EFT operators (initially only one or two), we can already extract a great deal of information on the Higgs-Goldstone interactions and hence, on the electroweak symmetry breaking (EWSB) mechanism.

At the energies currently reached by accelerators, both HEFT and SMEFT seem to fit the available observations. Recently, several works have addressed the distinction between the two EFTs, see for example~\cite{Brivio:2013pma, Gavela:2016vte, Cohen:2020xca, Alonso:2015fsp, Alonso:2016btr, Alonso:2016oah, Alonso:2021rac, Englert:2023uug, Bhardwaj:2023ufl, Ellis:2023zim, Liu:2023jbq, Graf:2022rco, deBlas:2018tjm, Pich:2015kwa, Arganda:2018ftn,Giudice:2007fh,Csaki:2015hcd}. Moreover, in previous works~\cite{Gomez-Ambrosio:2022qsi,Gomez-Ambrosio:2022why,Salas-Bernardez:2022hqv}, we showed that, from the SMEFT point of view, there is essentially one operator of the Warsaw basis that can, alone, shed light on the open questions of the symmetry breaking mechanism (this operator coincidentally being one of the least constrained by current LHC fits \cite{Englert:2015hrx, Ellis:2020unq, Ethier:2021bye, ATL-PHYS-PUB-2022-037} due to the narrow width approximation used by the experiments).  Furthermore, we showed that by comparing different scattering amplitudes of longitudinal gauge bosons one can draw a clear distinction between a SMEFT-like and a HEFT-like scenario. While current bounds on the $\omega\omega h$ and $\omega\omega hh$ interaction vertices are still loose, we found that this method could be useful in the future to eventually falsify the SMEFT approach (and hence the traditional EWSB realization 
{\it \`a la} SM).    

In this work, we go one step further by calculating analytical and numerical predictions for different cross sections. We focus on the electroweak production of 2, 3, and 4 Higgs bosons, which we study in various benchmark scenarios. In particular, we study the longitudinal vector boson scattering (VBS) into multiple Higgs bosons,    
which has been subject of study of several recent works, for example~\cite{Abouabid:2021yvw,Barman:2020ulr,Arganda:2018ftn,Basler:2019nas,Dawson:2017jja,Alasfar:2023xpc,Khoze:2017tjt,Khoze:2017uga}.

For the comparison of the SMEFT scenarios and the HEFT one, it will be fundamental to introduce the \textit{flare} function, $\mF(h)$. This function takes the role of parametrizing the Higgs-gauge interaction and it has the advantage that can be defined for both the SMEFT and HEFT theories\footnote{We will work in the energy regime where the equivalence theorem can be applied, identifying longitudinal $W$-bosons with scalar Goldstone bosons~\cite{Veltman:1989ud,Dobado:1993dg,Cornwall:1974km,Vayonakis:1976vz,Lee:1977eg,Chanowitz:1985hj,Pal:1994jk}.}, allowing us to map individual measurements to both scenarios. The flare function will be discussed in detail below (see also ref.~\cite{Gomez-Ambrosio:2022qsi}) and provides the effective coupling $a_n$ of a pair of EW gauge bosons, $WW$, to an arbitrary number $n$ of Higgs bosons: 
\begin{equation}
    \mF(h) = 1 + a_1 \frac{h}{v} + a_2 \left( \frac{h}{v} \right)^2 + a_3  \left( \frac{h}{v} \right)^3 + a_4 \left( \frac{h}{v} \right)^4 + \dots 
\end{equation}
It is customary that the first coefficients are also denoted by $a\equiv a_1/2$ and $b\equiv a_2$. In order to compute the production of Higgs bosons from the scattering of longitudinally polarized weak bosons, $W_LW_L\to n \times h$, we will employ the equivalence theorem (EqTh)~\cite{Veltman:1989ud,Dobado:1993dg,Cornwall:1974km,Vayonakis:1976vz,Lee:1977eg,Chanowitz:1985hj,Pal:1994jk}, where the EW gauge boson scattering $W_LW_L\to n \times h$ can be approximated by the EW Goldstone scattering $\omega\omega \to n \times h$ for $s\gg m_W^2$. Furthermore, both the EW gauge boson and Higgs boson masses will be neglected with respect to the center of mass energy of the process $\sqrt{s}$. Hence, we emphasize that the present article will be fully focused in the computation of these scattering amplitudes, $\omega\omega\to n \times h$, in the massless limit. More precisely, we will explore the production of two, three and four Higgs bosons, which we consider that are representative enough of the issue at hand.

Thus, this function $\mF(h)$ is going to provide the irreducible tree-level vertices for $\omega\omega\to n \times h$. However, in general, several couplings $a_j$ are going to contribute to the tree-level amplitude $\omega\omega\to n \times h$. All the amplitude and cross section calculations in this article will use this standard HEFT Lagrangian~\cite{Appelquist:1980vg,Longhitano:1980iz,Longhitano:1980tm,Feruglio:1992wf,Dobado:1990zh,Grinstein:2007iv,Shun-Zhi:2008oyf,LHCHiggsCrossSectionWorkingGroup:2016ypw} and the associated flare function $\mF(h)$~\cite{Gomez-Ambrosio:2022qsi,Gomez-Ambrosio:2022why,Salas-Bernardez:2022hqv}.

We display compact analytical expressions for the processes at hand. These are fundamental to understand patterns and simplifications that happen in the New Physics scenarios. For example, 
we find a factorization of the $s$-dependence in our analytical VBS expressions, such that only two numerical integrations are required to evaluate the cross sections at any energy, vastly reducing the computational cost. 
When looking at even higher Higgs multiplicities and full collider simulations this feature may be crucial.

Still, a superficial observation of these combinations of $a_j$ effective couplings may fail to reveal a pattern in the amplitudes. In fact, it is possible to eliminate the $h\omega\omega$ vertex by means of appropriate field redefinitions (see appendix~\ref{app:field-redef}). Although this distorts the effective Lagrangian and makes its symmetries less evident, this simplification makes the massless $\omega\omega\to n \times h$ computations much more transparent: the number of tree-level topologies is greatly reduced, with the diagrams now provided by an effective flare function with the form  
$    \hat{\mF}(h) = 1 \,  +\,  \hat{a}_2 \left( \frac{h}{v} \right)^2 + \hat{a}_3  \left( \frac{h}{v} \right)^3 +  \hat{a}_4 \left( \frac{h}{v} \right)^4 + \mO(h^5)$,       
where $\hat{a}_2=a_2-a_1^2/4$,  $\hat{a}_3 = a_3- \frac{2}{3}a_1 \left(a_2 -  a_1^2/4\right)$ and $\hat{a}_4 =  a_4 -   \frac{3}{4} a_1 a_3    +   \frac{5}{12}a_1^2\left(a_2-a_1^2/4\right)$. 
Further effective combinations are of no interest for this work although they can be extracted in a similar way. We have employed these simplified interactions to double-check all the computations. The technical details of this additional test have been relegated to appendix~\ref{app:field-redef}.

By studying longitudinal VBS into multiple Higgs bosons, this article explores the delicate cancellations one finds in these scattering amplitudes in the case when the low-energy range can be described by SMEFT. These cancellations are absent in more general theories that do not accept a low-energy EFT in terms of a Higgs doublet, leading to cross sections order of magnitude larger than those one obtain in SMEFT analyses, as we will see later in the main text.

In section~\ref{sec:HEFT-scat}, we compute the scattering amplitudes and cross sections for $\omega\omega$ into two, three and four Higgs bosons. Section~\ref{sec:SMEFT-scat} particularizes these outcomes to the case when the underlying theory accepts a SMEFT description at low energies. 
Section~\ref{sec:pheno} contains a phenomenological study of the obtained effective theory amplitudes. We use some benchmarks to compare both effective approaches, HEFT and SMEFT, showing how SMEFT models in general provide much more suppressed cross sections than theories which does not accept a SMEFT description. 
We show in section~\ref{sec:SMEFT-exclusion} how stringent this suppression must be, providing cross section exclusion plots for scenarios that admit SMEFT.  
Our final conclusions are provided in section~\ref{sec:conclusions}. Some technical aspects have been relegated to the appendices. 
All the results in this article have passed several checks: 
we have used an specific \textsc{Mathematica} code for the generation of arbitrary $\omega\omega\to n\times h$ amplitudes~\cite{AmplitudeCalculator}; we have used a HEFT model file~\cite{Martinez-Martin:MultiHiggsHEFT} in \textsc{FeynRules}~\cite{Rules} and \textsc{FeynCalc}~\cite{Calc} for the analytical calculations of the VBS amplitudes; the massless phase-space integration code {\tt MaMuPaXS}~\cite{MaMuPaXS} has been developed for the numerical evaluation of the $\omega\omega\to n\times h$ cross sections in the approximations of this article; finally, we have used the HEFT model file~\cite{Martinez-Martin:2022loy} in combination with {\tt MadGraph5\_aMC}~\cite{Alwall:2011uj} for a numerical test of the analytical expressions.

\section{Calculation in HEFT}
\label{sec:HEFT-scat}

In~\cite{Gomez-Ambrosio:2022qsi} we presented scattering amplitudes for different multiplicities of Higgs bosons in the final state. Here, we will focus on the cancellations occurring for SMEFT-like theories, in contrast to a generic HEFT scenario. 
We will be using the derivative terms of the leading order (LO) HEFT Lagrangian~\cite{Appelquist:1980vg,Longhitano:1980iz,Longhitano:1980tm,Feruglio:1992wf,Dobado:1990zh,Grinstein:2007iv,Shun-Zhi:2008oyf,LHCHiggsCrossSectionWorkingGroup:2016ypw},  
\begin{eqnarray}
\mL_{\rm HEFT} &=& \frac{1}{2}(\partial_\mu h)^2 + \frac{v^2}{4}\mF(h)\, \mbox{Tr}\left\{\partial_\mu U^\dagger \partial^\mu U\right\} \, ,
\label{eq:HEFT-Lagr}
\end{eqnarray}
where the unitary matrix $U(\omega)=1+i\sigma^a \omega^a/v+\mO(\omega^2)$ parametrizes the $SU(2)_L\times SU(2)_R/SU(2)_{L+R}$ coset in terms of the EW Goldstone bosons $\omega^a$. A usual representation is provided, e.g., by the exponential form $U=\exp\{ i\omega^a \sigma^a/v\}$. Nevertheless, for the tree-level $\omega\omega\to n\times h$ scattering discussed in this article, one only needs the $\mO(\omega)$ term in $U$, which is identical in all representations. 
 Finally, we will neglect next-to-leading order, $\mO(p^4)$, HEFT contributions~\cite{LHCHiggsCrossSectionWorkingGroup:2016ypw,Alonso:2012px,Buchalla:2013rka,Krause:2018cwe,Pich:2018ltt,Herrero:1993nc,Pich:2016lew,Buchalla:2012qq}, both at tree and one-loop level.

For this work, eq.~(\ref{eq:HEFT-Lagr}) provides the relevant part of the HEFT Lagrangian because we are working under the following set of restrictions: 
First, we are only considering the derivative part of the lowest order chiral Lagrangian, $\mO(\partial^2)$, as we will be choosing a kinematic regime well over the production threshold ($s\gg  m_W^2\sim m_h^2$); Second, eq.~(\ref{eq:HEFT-Lagr}) incorporates only the scalar sector, as we will be using the equivalence theorem to compute the amplitudes $W_LW_L\to n \times h$ at tree-level for $s\gg m_W^2$~\cite{Veltman:1989ud,Dobado:1993dg,Cornwall:1974km,Vayonakis:1976vz,Lee:1977eg,Chanowitz:1985hj,Pal:1994jk}. In this article we will focus on the charged Goldstone ($\omega^\pm$) scattering, although a similar analysis could be done with the neutral ones ($\omega^0$). This exercise can be repeated similarly for the Yukawa sector through the study of $t\bar{t}\to n\times h$ processes, where one would also need to introduce the Yukawa flare function, which parametrizes the relevant local vertices (see~\cite{Gomez-Ambrosio:2022why,Bhardwaj:2023ufl}).

Thus, we will compute the $\, \omega^+\omega^-\to n\times h \,$ amplitudes, 
 $T_{\omega\omega\to n \times h}$, and analyse the corresponding cross section, 
\begin{equation}
\sigma_{\omega\omega\to n \times h} \,=\, \frac{1}{n!} \, \frac{1}{2 s} \, \displaystyle{\int\lvert T_{\omega\omega\to n \times h}\rvert^2 \, d\Pi_n } \, .    
\label{eq:Tcross-section}
\end{equation}
It is useful to note that, in the massless approximation, one can fully factor $s$ out of the phase-space integration  $d\Pi_n= s^{n-2} d\widetilde{\Pi}_n$. In this way we are left with a pure angular integration in $d\widetilde{\Pi}_n$. Further details like, e.g., the analytical expression for the total phase-space volume $\mV_n=\int d\Pi_n$, can be found in appendix~\ref{app:phase-space}.

In the following subsections we describe the amplitudes and cross sections for the most general case, that of HEFT. In  section~\ref{sec:SMEFT-scat} we will be restricting them to the SMEFT case.

\subsection{\texorpdfstring{$\omega\omega\to 2h$}{ww->2h}}

In the simplest case, $\omega^+\omega^-\to 2h$, one finds in HEFT~\cite{DelgadoLopez:2016cty, Delgado:2013loa,Domenech:2022uud}: 
\begin{equation}
T_{\omega\omega\to 2h} 
\,=\, -\, \frac{\hat{a}_2   s}{v^2} \,  ,
\label{eq:Thh-HEFT}
\end{equation}
with $\hat{a}_2=a_2-a_1^2/4 = b-a^2$. This result is in agreement with previous effective theory $WW\to 2h$ studies~\cite{Anisha:2022ctm,Arganda:2018ftn,Domenech:2022uud,Dobado:2017lwg} at high energies, where the Equivalence Theorem approximation is applicable. 

The amplitude in eq.~(\ref{eq:Thh-HEFT}) is a pure $J=0$ wave, depending only on the total center-of-mass (CM) energy, not on the scattering angle. Thus, one can readily compute the total cross section,  
\begin{equation}
\sigma_{\omega\omega\to 2h} 
\,=\,  \frac{8\pi^3 \, \hat{a}_2^2 }{s}\, \left( \frac{s}{16\pi^2 v^2}\right)^2\, \, .
\end{equation}

\subsection{\texorpdfstring{$\omega\omega\to 3h$}{ww->3h}}

For the $\omega\omega \to 3h$ amplitude, we have checked our result with the aid of \textsc{FeynRules} and \textsc{FeynCalc} (based on the previous work in~\cite{Martinez-Martin:MultiHiggsHEFT,Martinez-Martin:2022loy}) and an independent \textsc{Mathematica} implementation by the authors~\cite{AmplitudeCalculator}, obtaining, 
\begin{eqnarray} \label{amp:3h}
    T_{\omega\omega\to 3h}
    &=&
    \,-\, \frac{3 \hat{a}_3 s}{v^3} \,,
\end{eqnarray}
with $\hat{a}_3 = a_3- \frac{2}{3}a_1 \left(a_2 -  a_1^2/4\right) = a_3 - \frac{4}{3} a \left(b-a^2\right)$.  
This amplitude readily leads us to the corresponding total cross section:
\begin{eqnarray}
\sigma_{\omega\omega\to 3h}&=&  
\frac{12\pi^3 \,\hat{a}_3^2} {s}\left(\frac{s}{16\pi^2 v^2}\right)^3\, .
\end{eqnarray} 

Previous works have analyzed the $WW\to 3h$ scattering in the context of effective theories~\cite{Gonzalez-Lopez:2020lpd,Chen:2021rid}, considering modifications in the $WWh$, $WWh^2$, $h^3$ and $h^4$ vertex couplings.   
One of the novelties of the present work is to incorporate the full generality of allowed $\omega\omega h^n$ effective vertices relevant for the process (and the general $WW h^n$ vertices beyond the EqTh). In particular, we considered the $\omega\omega h^3$ coupling ($a_3$). Nonetheless, we find that at high energies its contribution to the cross section yields the same energy and angular structure as those from $a_1$ and $a_2$, being the relevant information encapsulated in the combination   $\hat{a}_3 = a_3- \frac{2}{3}a_1 \left(a_2 -  a_1^2/4\right)$.

\subsection{\texorpdfstring{$\omega\omega\to 4h$}{ww->4h}}

In this subsection we compute the amplitude of $\omega^+(k_1)\, \omega^-(k_2) \to h(p_1)\, h(p_2)\, h(p_3)\, h(p_4)$ in HEFT in the same way as in the previous cases. 
This amplitude can be written as
\begin{eqnarray} 
    T_{\omega\omega\to 4h} &=&
  \, -\, \frac{4 s}{v^4}\left(3\hat{a}_4 + \hat{a}_2^2 \, (B-1)  \right)        \, ,      
\label{eq:4h-HEFT-amp}
\end{eqnarray}  
where $\hat{a}_2=a_2-a_1^2/4=b-a^2 $ and  
$\hat{a}_4 =  a_4 -   \frac{3}{4} a_1 a_3 + \frac{5}{12}a_1^2\left(a_2-a_1^2/4\right) = a_4 - \frac{3}{2} a\,a_3 + \frac{5}{3}a^2\left(b-a^2\right)$.

The dimensionless kinematic function $B$ can be written as
\begin{eqnarray} 
\label{eq:B}
B &=& f_1f_2f_3f_4 \,\bigg(\mathcal{B}_{1234} + \mathcal{B}_{1324} + \mathcal{B}_{1423} + \mathcal{B}_{2314} + \mathcal{B}_{2413} + \mathcal{B}_{3412}\bigg) \, ,
\end{eqnarray}
with 
\begin{eqnarray}
\mathcal{B}_{ijk\ell} \,=\, \frac{z_{ij}z_{k\ell}}{2f_if_jz_{ij}-f_iz_i-f_jz_j}\, ,
\end{eqnarray}  
where $f_i = q p_i/q^2$, $z_i = 2 k_1 p_i/q p_i$, $z_{ij} = z_{ji}= q^2\, (p_ip_j)/[(q p_i)\, (q p_j)]$ and the total four-momentum $q=k_1+k_2=p_1+p_2+p_3+p_4$. 
In the centre-of-mass (CM) rest-frame these relations define: 
the three-momentum fractions $f_i= \lVert\vec{p}_i\rVert/\sqrt{s}$ ($s=4\lVert\vec{k}_1\rVert^2$) for each outgoing Higgs boson; 
the angular functions $z_i= 2\sin^2(\theta_i/2)$ with $\theta_i$ being the angle between the $i$-th Higgs boson and the incoming $\omega^+$ 
Goldstone boson momenta, $\vec{k}_1$  
(that is, $z_1=1-\cos\theta$, $z_2=1+\cos\theta$ as usual in a two-body problem with $t$ and $u$ channels);  
$z_{ij}= 2\sin^2(\theta_{ij}/2)$,  with $\theta_{ij}$ being the angle between the $i$-th and $j$-th Higgs bosons.   
Total 4-momentum conservation allows us to establish some relations between these parameters which are used to simplify the expression for the amplitude. More details are given in appendix~\ref{app:mom-conservation}, where we also present $B$ in terms of $f_{1,2,3}$, $z_{1,2,3}$,  $z_{12}$ and $z_{13}$, reducing the number of free kinematic variables.

\begin{figure}[!t] 
     \centering
     \qquad
     \begin{subfigure}[]
         \centering
         \begin{tikzpicture}[scale=1]
     \draw[dashed] (-1,1) -- (0,0)-- (-1,-1);
     \draw[] (0,0)-- (1,1);
     \draw[] (0,0)-- (1,0.35);
     \draw[] (0,-0)-- (1,-0.35);
       \draw[] (0,0)-- (1,-1);
       \draw[] (-1.25,1) node {$\omega$};
       \draw[] (-1.25,-1) node {$\omega$};
     \draw[] (1.25,1) node {$h$};
     \draw[] (1.25,-0.35) node {$h$};
     \draw[] (1.25,0.35) node {$h$};
          \draw[] (1.25,-1) node {$h$}; 
\end{tikzpicture}
     \end{subfigure}
     \qquad
          \begin{subfigure}[]
         \centering
         \begin{tikzpicture}[scale=1]
     \draw[dashed] (-1,1) -- (0,1)--(0,-1)-- (-1,-1);
     \draw[] (0,1)-- (1,1);
     \draw[] (0,1)-- (1,0.35);
     \draw[] (0,1)-- (1,-0.35);
       \draw[] (0,-1)-- (1,-1);
       \draw[] (-1.25,1) node {$\omega$};
       \draw[] (-1.25,-1) node {$\omega$};
     \draw[] (1.25,1) node {$h$};
     \draw[] (1.25,-0.35) node {$h$};
     \draw[] (1.25,0.35) node {$h$};
          \draw[] (1.25,-1) node {$h$}; 
\end{tikzpicture}
     \end{subfigure}
          \qquad
     \begin{subfigure}[]
         \centering
         \begin{tikzpicture}[scale=1]
     \draw[dashed] (-1,1) -- (0,1)--(0,-1)-- (-1,-1);
     \draw[] (0,1)-- (1,1);
     \draw[] (0,0)-- (1,0.35);
     \draw[] (0,0)-- (1,-0.35);
       \draw[] (0,-1)-- (1,-1);
       \draw[] (-1.25,1) node {$\omega$};
       \draw[] (-1.25,-1) node {$\omega$};
     \draw[] (1.25,1) node {$h$};
     \draw[] (1.25,-0.35) node {$h$};
     \draw[] (1.25,0.35) node {$h$};
          \draw[] (1.25,-1) node {$h$}; 
\end{tikzpicture}
     \end{subfigure}
    
     \qquad
     \begin{subfigure}[]
         \centering
         \begin{tikzpicture}[scale=1]
     \draw[dashed] (-1,1) -- (0,1)--(0,-1)-- (-1,-1);
     \draw[] (0,1)-- (1,1);
     \draw[] (0,1)-- (1,0.35);
     \draw[] (0,-1)-- (1,-0.35);
       \draw[] (0,-1)-- (1,-1);
       \draw[] (-1.25,1) node {$\omega$};
       \draw[] (-1.25,-1) node {$\omega$};
     \draw[] (1.25,1) node {$h$};
     \draw[] (1.25,-0.35) node {$h$};
     \draw[] (1.25,0.35) node {$h$};
          \draw[] (1.25,-1) node {$h$}; 
\end{tikzpicture}
     \end{subfigure}
    \qquad
     \begin{subfigure}[]
         \centering
         \begin{tikzpicture}[scale=1]
     \draw[dashed] (-1,1) -- (0,1)--(0,-1)-- (-1,-1);
     \draw[] (0,1)-- (1,1);
     \draw[] (0,1)-- (1,0.35);
     \draw[] (0,-.35)-- (1,-0.35);
       \draw[] (0,-1)-- (1,-1);
       \draw[] (-1.25,1) node {$\omega$};
       \draw[] (-1.25,-1) node {$\omega$};
     \draw[] (1.25,1) node {$h$};
     \draw[] (1.25,-0.35) node {$h$};
     \draw[] (1.25,0.35) node {$h$};
          \draw[] (1.25,-1) node {$h$}; 
\end{tikzpicture}
     \end{subfigure}
     \qquad
     \begin{subfigure}[]
         \centering
         \begin{tikzpicture}[scale=1]
     \draw[dashed] (-1,1) -- (0,1)--(0,-1)-- (-1,-1);
     \draw[] (0,1)-- (1,1);
     \draw[] (0,0.35)-- (1,0.35);
     \draw[] (0,-0.35)-- (1,-0.35);
       \draw[] (0,-1)-- (1,-1);
       \draw[] (-1.25,1) node {$\omega$};
       \draw[] (-1.25,-1) node {$\omega$};
     \draw[] (1.25,1) node {$h$};
     \draw[] (1.25,-0.35) node {$h$};
     \draw[] (1.25,0.35) node {$h$};
          \draw[] (1.25,-1) node {$h$}; 
\end{tikzpicture}
     \end{subfigure}
      \qquad

   \caption{
   {\small  Relevant tree-level topologies for $\omega\omega\to 4h$. 
   Note that in addition one needs to consider all possible permutations for the assignment of the external particles.
   }}
   \label{fig:4h-diagrams}  
\end{figure}
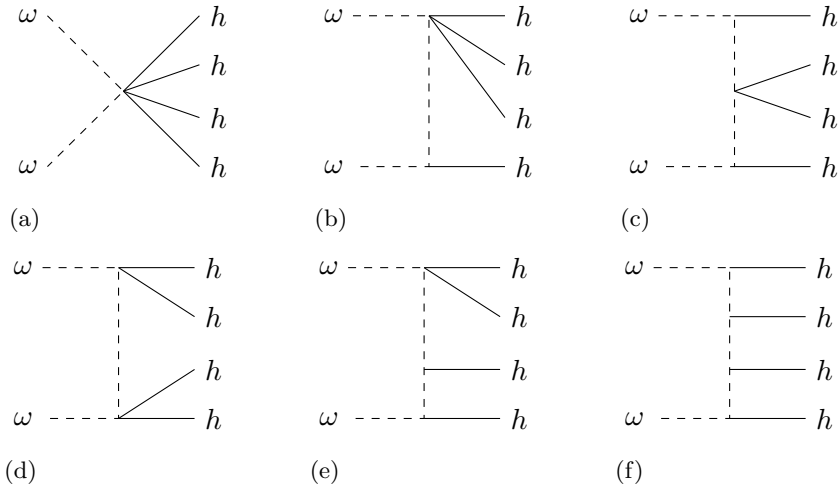

The term with $B$ provides the irreducible contribution from cross-propagator exchanges (see appendix \ref{app:field-redef}). For the production of 2 and 3 Higgs bosons, even though we had cross-channel exchanges, the propagator poles always cancelled with the vertices when the external legs were set on-shell. This is not fully possible in the 4 Higgs case and the $B$ term remains.

Figure~\ref{fig:4h-diagrams} shows the tree-level diagrams contributing to the $\omega\omega\to 4 h$ process.  
From the structure of the various possible tree-level diagrams, containing crossed-channel $\omega$ exchanges, one might think that the amplitude~(\ref{eq:4h-HEFT-amp}) may become singular for some points of the phase-space integration. Indeed, there are diagrams that contain up to three intermediate $\omega$ propagators. Nonetheless, a careful observation shows that the momentum-dependent vertex structure exactly cancels the propagators in several cases. 
 
Diagram {\bf a)} has no propagator. Likewise, when the external legs are set on-shell, the intermediate propagators in diagram  {\bf b)} and {\bf c)} cancel out --diagram by diagram-- and their contribution to the amplitude is simply a polynomial in momenta. 
However, for this process we find that there are topologies where one of the intermediate propagators survive, {\bf d)}, {\bf e)}, and {\bf f)}, generating the $B$ denominators which one finds in~(\ref{eq:4h-HEFT-amp}) and~(\ref{eq:B}).  

One might think that the surviving propagator in diagrams {\bf d)}, {\bf e)} and {\bf f)} yields poles in the amplitude at certain energies. However, these potential singularities are
always proportional to the structure $\mM= \, -4 C (k_1 \ell)(k_2 \ell)/\ell^2= C (T-M_A^2)(T-M_B^2)/T$, with $P_A$ and $P_B$ the four-momenta of two disjoint pairs of final Higgs bosons (e.g., $P_A=p_1+p_2$ and $P_B=p_3+p_4$), $\ell= k_1-P_A=P_B-k_2$, $T=\ell^2$, $M_{A,B}=\sqrt{P_{A,B}^2}$ and $C$ a constant proportional to $v^{-4}$. In the physical region where $\sqrt{s}\geq M_A+M_B$ one has $T<0$ when both $M_A>0$ and $M_B>0$. Thus, the amplitude never turns singular in the phase-space integral as the singularity is located at non-physical energies\footnote{
For $k_1^2=k_2^2=0$ one has $T_{\rm max\, (min)} = \left[-\Delta s  \pm\sqrt{(\Delta s)^2 - 4 M_A^2 M_B^2} \right]/2$ (with $\Delta s=s-M_A^2-M_B^2$), where $T_{\rm min}\leq T\leq T_{\rm max}$. For $M_A>0$ and $M_B>0$, $\Delta s\geq 2 M_A M_B>0$ in the physical region, leading to strictly negative values of $T_{\rm max\, (min)}$ and $T$.}.   
Furthermore, when either $M_A$ or $M_B$ are zero, the diagram becomes non-singular in the whole complex plane. In summary, as it happened with the two and three Higgs final states, the $\omega\omega\to 4h$ cross section is infrared safe and does not contain pole  singularities.

This amplitude provides the total cross section for $\omega\omega\to 4h$:  
\begin{eqnarray}
\sigma_{\omega\omega\to 4h} &=&  \frac{8\pi^3}{9 s}   \left(\frac{s}{16\pi^2 v^2}\right)^4  
\left[\left(3\hat{a}_4 - \hat{a}_2^2 \right)^2   + 2 \left(3\hat{a}_4 - \hat{a}_2^2 \right) \hat{a}_2^2 \chi_1 +\hat{a}_2^4\chi_2\right] \, ,
\end{eqnarray}
with the normalized phase-space integrals,  
\begin{eqnarray}
&&  \chi_n =\mathcal{V}_4^{-1}\int d\Pi_4 \,  B^n  \, , 
\label{eq:chi-n_def}
\end{eqnarray} 
and the total 4-particle phase-space volume $\mathcal{V}_4= {\int d\Pi_4}=s^2 \left(24 (4\pi)^5 \right)^{-1}$, rendering the $\chi_n$ dimensionless and $s$-independent (see appendix~\ref{app:phase-space} for further details). 
These two constants are dimensionless and independent of the effective Lagrangian parameters. We evaluate them numerically through our phase-space integration code ({\tt MaMuPaXS})~\cite{MaMuPaXS}, obtaining: 
\begin{eqnarray}
\chi_1 =
 -0.124984\, (10)\,, \qquad   
\chi_2 = 
0.0193760 \, (16)\, .
\label{eq:chi-n_num}
\end{eqnarray}

The angular function $B$ is given in eq.~\eqref{eq:B} and its value is independent of the Higgs parameters. Notice that $\chi_1$ and $\chi_2$ are defined in the massless limit, therefore $s$-independent.  
Due to the normalization factor $\mathcal{V}_4^{-1}$ in~(\ref{eq:chi-n_def}), the $\chi_j$ are pure angular integrations. 
The total phase-space volume is given by $\mathcal{V}_n=\int d\Pi_n  =   s^{n-2} \left(  2(4\pi)^{2n-3} \Gamma(n)\Gamma(n-1)\right)^{-1}$ 
(see appendix~\ref{app:phase-space} for further details).  
Finally note that both type of diagrams, either with $\hat{a}_4$ or $\hat{a}_2^2$, have a similar weight in the cross section; the $\hat{a}_2^2$ terms appear multiplied by $(B-1)$ in the amplitude and hence lead to the integrals $\mathcal{V}_4^{-1}\int d\Pi_4 \, (B-1)^n \sim \mO(1)$ in the cross section. For convenience, we have explicitly separated the $B^n$ integrals, which need to be computed numerically. 

We have cross-checked these analytical results for $\omega\omega\to 2h$, 
$\omega\omega\to 3h$ and $\omega\omega\to 4h$ with {\tt MadGraph5\_aMC}, finding a complete agreement with its numerical determinations.  
In particular, we have also used the latter to check the numerical values for $\chi_1$ and $\chi_2$ obtained with {\tt MaMuPaXS}~\cite{MaMuPaXS}. 

\section{SMEFT treatment: suppression of multi-Higgs production}
\label{sec:SMEFT-scat}

In a previous work~\cite{Gomez-Ambrosio:2022qsi} we discussed that, neglecting custodial symmetry breaking terms, 
there is only one operator that modifies the flare function $\mF(h)$ at each perturbative order in the SMEFT expansion. As it is also done for dimension $D=6$ in the Warsaw basis~\cite{Grzadkowski:2010es}, one can use equations of motion and field redefinitions to remove SMEFT operators proportional to $\Box H$ (with $H$ the Higgs doublet). We can choose an EFT operator basis that optimizes the present analysis so only one operator is relevant for the flare function at each $D$ dimension. 
Hence, for theories that admit a SMEFT description one has, up to $\mO(1/\Lambda^4)$, explicit expressions for the coefficients of $\mathcal{F}(h)$, in terms of the SMEFT Wilson coefficients and scale $\Lambda$:
\begin{eqnarray}
a_1/2\, &=& \, a \, = \, 1 
\, +\,  \frac{d}{2}\, +\, \frac{d^2}{2}\left(\frac{3}{4} +\rho\right)      \,+\,\mathcal{O}\left(d^3\right)\, , \nonumber\\
a_2 \, &=& \, b \, = \, 1
\, +\,  2d  \, +\, 3 d^2\left(1+\rho\right) \,+\,\mathcal{O}\left(d^3\right)\, , 
\nonumber\\
a_3 \, &=&  \frac{4}{3}d  \, +\,  d^2\left(\frac{14}{3}+4\rho\right) \,+\,\mathcal{O}\left(d^3\right)\, ,  
\nonumber\\
a_4 \, &=&  \frac{1}{3}d  \, +\,  d^2\left(\frac{11}{3}+3\rho\right) \,+\,\mathcal{O}\left(d^3\right)\, ,  
\nonumber\\
a_5 \, &=&    d^2\left(\frac{22}{15}+\frac{6}{5}\rho\right) \,+\,\mathcal{O}\left(d^3\right)\, ,  
\nonumber\\
a_6 \, &=& d^2\left(\frac{11}{45}+ \frac{1}{5}\rho\right) \,+\,\mathcal{O}\left(d^3\right)\, ,
\label{eq:SMEFT-FF}
\end{eqnarray}
with,
\begin{equation}
    d=\frac{2v^2 c_{H\Box}^{(6)}}{\Lambda^2} \,,  \qquad \rho=\frac{c_{H\Box}^{(8)}}{2 (c_{H\Box}^{(6)})^2}\;,
    \label{eq:d-rho-defs}
\end{equation}
where $c_{H\Box}^{(D)}$ is the dimensionless Wilson coefficient for the operator $|H|^{D-4} \Box |H|^2/\Lambda^{D-4}$. Note that the expressions in~(\ref{eq:SMEFT-FF}) are truncated at $\mO(d^2)=\mO(v^4/\Lambda^4)$, and higher powers of $d$ have been neglected. Higher coefficients $a_n$ with $n\geq 7$ vanish at this order in SMEFT.

From~(\ref{eq:SMEFT-FF}) and section~\ref{sec:HEFT-scat} results we can read the relevant combinations of $a_j$ couplings that determine the $\omega\omega$ scattering into two, three and four Higgs bosons in SMEFT:
\begin{eqnarray}
\hat{a}_2 &=& d\, +\, 2 d^2 (1+\rho)\, +\, \mO(d^3)\, ,\nonumber\\
\hat{a}_3 &=& \frac{4}{3} d^2 (1+\rho)\, +\, \mO(d^3)\, ,\nonumber\\
\hat{a}_4 &=& \frac{1}{3} d^2  (  2  
+\rho)\, +\, \mO(d^3)\, . 
\label{eq:SMEFT-ajhat}
\end{eqnarray} 
We note that these combinations are also found after considering the field redefinitions of appendix~\ref{app:field-redef}.

It is not difficult to observe that for $d\neq 0$ these expressions obey the $\mO(d)\sim \mO(1/\Lambda^2)$ SMEFT relations~\cite{Gomez-Ambrosio:2022why,Salas-Bernardez:2022hqv}, 
\begin{equation}
\Delta a_1 \,=\,  \frac{1}{2}\Delta a_2 \,=\, \frac{3}{4} a_3 \,=\, 3 a_4\, , \qquad 
a_{k\geq 5}=0\, ,
\end{equation}
with $\Delta a_1\equiv a_1-2$ and $\Delta a_2\equiv a_2-1$. These SMEFT relations have been refined and positively tested up to  $\mO(d^2)\sim \mO(1/\Lambda^4)$. 

On the other hand, it is worth mentioning here that some UV-completions, such as the 2-Higgs Doublet Model, lead to a SMEFT low-energy theory but do not contribute to $\mF(h)$ at $\mO(1/\Lambda^2)$~\cite{Dawson:2022cmu,Dawson:2023ebe} (i.e., $c_{H\Box}^{(6)}$ and $d$ vanish). Their first contribution to the flare function appears at $\mO(1/\Lambda^4)$ through a non-zero $c_{H\Box}^{(8)}$ Wilson coefficient (this is, through a non-zero $d^2\rho$). In this case, the SMEFT relations in~\cite{Gomez-Ambrosio:2022why,Salas-Bernardez:2022hqv} get modified~\cite{Dawson:2023ebe} into the $\mO(1/\Lambda^4)$ constraints, 
\begin{equation}
\Delta a_1 \,=\,  \frac{1}{3}\Delta a_2 \,=\, \frac{1}{4} a_3 \,=\, \frac{1}{3} a_4\,=\, \frac{5}{6} a_5 \,=\, 5 a_6\, , \qquad 
a_{n\geq 7}=0\, ,
\end{equation}
with $\Delta a_1= 2 \Delta a= 2 c_{H\Box}^{(8)} v^4/\Lambda^4 =d^2 \rho\neq 0$.

\subsection{\texorpdfstring{$\omega\omega\to 2h$}{ww->2h}}
\label{sec:SMEFT-2h}

Replacing the expressions of eq.~(\ref{eq:SMEFT-FF}) in~(\ref{eq:Thh-HEFT}) and expanding in  powers of $1/\Lambda^2$ one obtains the SMEFT prediction for the $\omega^+\omega^-\to 2h$ scattering amplitude: 
\begin{eqnarray} 
T_{\omega\omega\to 2h} &=& 
\, -\, \frac{s}{v^2}
\, \left[   d\, + \, 2 d^2\, (1+\rho)\right]\,+\,\mathcal{O}\left(d^3\right)  \, ,
\\ \label{cs2}
\sigma_{\omega\omega\to 2h} &=& 
\frac{8\pi^3}{s} \left[ d^2 \,+\, 4 d^3\, \left(1+\rho\right)\right] \, \left(\frac{s}{16\pi^2 v^2}\right)^2\,+\,\mathcal{O}\left(d^4\right)\, .  
\end{eqnarray}  
One can read from these expressions that, for the $\omega\omega\to 2h$ amplitude and cross section, the LO contributions in the SMEFT expansion are $\mO(d)\sim \mO(1/\Lambda^2)$ and  $\mO(d^2)\sim \mO(1/\Lambda^4)$, respectively. These terms are produced by the dimension $D=6$ SMEFT Lagrangian. 
However, since we are also discussing the impact from the dimension $D=8$ SMEFT operators, we also provide the next corrections to the amplitude and cross section ($\mO(d^2)\sim \mO(1/\Lambda^4)$ and  $\mO(d^3)\sim \mO(1/\Lambda^6)$), which are fully determined by the $D=6$ and $D=8$ Lagrangians. 

\subsection{\texorpdfstring{$\omega\omega\to 3h$}{ww->3h}}
\label{sec:SMEFT-3h}

For $n=3$ Higgs bosons in the final state, we have the scattering amplitude and cross section,
\begin{eqnarray}
T_{\omega\omega\to 3h}&=& 
-\frac{4 s}{v^3}d^2\left(1+ \rho \right) \,+\,\mathcal{O}\left(d^3\right)  \, ,
\\ \label{cs3}
\sigma_{\omega\omega\to 3h}&=& 
\frac{64\pi^3}{ 3s}   \, d^4\, (1+\rho)^2 \,  
\left(\frac{s}{16\pi^2 v^2}\right)^3  \,+\,\mathcal{O}\left(d^5\right)\, . 
\label{amp_s_3h}
\end{eqnarray}
We are providing the three-Higgs amplitude and cross section at the lowest non-trivial order in the SMEFT expansion, $\mO(d^2)\sim \mO(1/\Lambda^4)$ and  $\mO(d^4)\sim \mO(1/\Lambda^8)$, respectively.

\subsection{\texorpdfstring{$\omega\omega\to 4h$}{ww->4h}}
\label{sec:SMEFT-4h}

For $n=4$ Higgs bosons in the final state, we have the scattering amplitude and cross section,
\begin{eqnarray}
T_{\omega\omega\to 4h}&=&-\frac{4s}{v^4}d^2\left(1+\rho+B\right) \,+\,\mathcal{O}\left(d^3\right)  \, ,   \\ \label{cs4}
\sigma_{\omega\omega\to 4h}&=&      
     \frac{8\pi^3}{9 s}\left(\frac{s}{16\pi^2 v^2}\right)^4  \, d^4\,  \bigg[ (1+\rho)^2 
+ 2(1+\rho)\chi_1  + \chi_2 
\bigg]\, +\, \mO(d^5)\,,
\end{eqnarray}
with the angle-dependent function $B$ from 
eq.~(\ref{eq:B}) and the numerical evaluation in $\chi_1$ and $\chi_2$ provided in eq.~(\ref{eq:chi-n_def}) through our phase-space integration code {\tt MaMuPaXs}~\cite{MaMuPaXS}. Again, we have computed the amplitude and cross section at their lowest non-trivial order in the SMEFT expansion, $\mO(d^2)\sim \mO(1/\Lambda^4)$ and  $\mO(d^4)\sim \mO(1/\Lambda^8)$, respectively.

\subsection{\texorpdfstring{$\omega\omega\to n \times h$}{ww->n x h}: multi-Higgs production suppression with \texorpdfstring{$\mL^{\rm SMEFT}_{(D=6)}$}{L SMEFT(D=6)}}

For 3 or 4 Higgs bosons in the final state we find that the amplitude exactly cancels at $\mO(1/\Lambda^2)$. The first contributions appear at $\mO(1/\Lambda^4)$ from diagrams with either two $\mO(1/\Lambda^2)$ vertices from operators in $\mL_{D=6}^{\rm SMEFT}$ or one $\mO(1/\Lambda^4)$ vertex from $\mL_{D=8}^{\rm SMEFT}$. 
 
At lowest non-vanishing order in the SMEFT expansion, for $\omega\omega\to n \times h$,  we find the SMEFT amplitude:
\begin{equation}
T_{\omega\omega\to n \times h} \, =   \, C_n(z;c_j)\, 
 \frac{s}{ v^{n} }  \, \left(\frac{v^2}{  \Lambda^{2}}\right)^{\gamma_n} \, ,   
\end{equation} 
up to  corrections with higher powers in $s$. The $C_n(z;c_j)$ are $\mO(1)$ dimensionless combination of angular functions the $z=\{z_i,z_{jk}\}$ trigonometric functions previously defined (see appendix~\ref{app:mom-conservation}) and the $\mO(1)$ SMEFT Wilson coefficients $c_j$ (where $\Lambda^{4-D}$ has been factored out from the SMEFT Lagrangian term as it is customary).  

The exponent $\gamma_n$ indicates how many insertions of the dimension-6 SMEFT operators one needs to introduce in the diagram to obtain the first non-vanishing contribution to $\omega\omega\to n \times h$. It follows the pattern $\gamma_2=1$, $\gamma_3=\gamma_4=2$, $\gamma_5=\gamma_6=3$... This can be summarized in the general expression $\gamma_n= \lceil n/2 \rceil$ (see below and appendix \ref{app:SMEFT-multiH-sup}). 

Up to $\mO(1/\Lambda^2)$, the SMEFT Lagrangian for the scalar sector is given by the operators~\cite{Gomez-Ambrosio:2022qsi}, 
\begin{equation}
\mL \,=\, \lvert\partial H\rvert^2 \, + \, \frac{c_{H\Box}^{(6)}}{\Lambda^2} \lvert H\rvert^2\Box \lvert H\rvert^2\, +\, \dots
\label{eq:LSMEFT-dominant}
\end{equation}
where the dots stand for non-derivative operators, and other SMEFT operators of higher canonical dimension.  

The first term in the r.h.s. comes from the dimension $D=4$ part of Lagrangian and the second one from $\mL^{\rm SMEFT}_{(D=6)}$.
At high enough energies over the production thresholds ($m\ll \sqrt{s}$) one may neglect the non-derivative terms and higher dimension operators will be subdominant for energies much lower than the effective theory cutoff ($\sqrt{s}\ll \Lambda$). 
Although this article has been focused on custodial preserving operators, custodial breaking terms would show a similar structure at $D=6$, with all the conclusions drawn in this subsection remaining unchanged if they were also incorporated.

For the reasoning in this subsection it will be better to handle the Higgs doublet in Cartesian coordinates, in such a way that $2\lvert H\rvert^2= (v+h)^2 + \sum_a(\omega^a)^2$.  Also, notice that we are considering a linear representation of $H$, different to the non-linear one used in the previous sections. 
Abusing notation, we have denoted both linear and non-linear Goldstone and Higgs fields as $h$ and $\omega^a$.   
Hence, the fields $h$ and $\omega^a$ in each representation should not be confused. Nonetheless, both representations of the Goldstone and Higgs fields lead to the same on-shell $S$-matrix elements. 

With this consideration in mind, the first term on the r.h.s of eq.~(\ref{eq:LSMEFT-dominant}) provides the kinetic term of the Higgs $h$ and EW Goldstones $\omega^a$. 

The second term on the r.h.s. of eq.~(\ref{eq:LSMEFT-dominant}) yields three types of contribution: vertices with 4 scalars, vertices with 3 scalars and a correction to the kinetic term (which will need to be rescaled to provide the canonically normalized propagator). Note that all these operators contain two derivatives. 

Feynman diagrams built by such Lagrangian will, in general, contain $N_{V_3}$ vertices of 3 legs, $N_{V_4}$ vertices of 4 legs, $N_I$ internal propagators, $N_E$ external lines and $L$ loops. Here we will set the number of loops to zero and the amplitudes $\omega\omega\to n \times h$ will have $N_E=n+2$, obtaining: 
\begin{equation}
N_{V_4}+\frac{1}{2}N_{V_3} \,=\, L \, +\, \frac{n}{2}\, \stackrel{L=0}{=} \, \frac{n}{2}\, ,
\label{eq:topo-eq}
\end{equation}
where we eliminated $N_I$ by means of the relations $3 N_{V_3} + 4 N_{V_4}= 2 N_I + N_E$ and 
$L-N_I+N_V=1$, with the total number of vertices $N_V=N_{V_3}+N_{V_4}$.

Tree-level diagrams will show a SMEFT $1/\Lambda^2$ scaling of the form, 
\begin{equation}
T_{\omega\omega\to n \times h}\,\sim\, \left(\frac{1}{\Lambda^2}\right)^{N_{V_3}+N_{V_4}}\, =\, \left(\frac{1}{\Lambda^2}\right)^{1+N_I}\, .
\end{equation}
Using $L-N_I +(N_{V_3}+N_{V_4})=1$ and eq.~(\ref{eq:topo-eq}), one obtains $1+N_I=N_{V_3}+N_{V_4}=\frac{1}{2}(n+N_{V_3})$ at tree-level. 
Thus, for a given number of final Higgses $n$, we find that 
the leading diagrams scales like,
\begin{equation}
T_{\omega\omega\to n \times h}\,\sim\, \left(\frac{1}{\Lambda^2}\right)^{\gamma_n}\, , 
\qquad \mbox{with }  \gamma_n\equiv 1+N_I^{\rm min}=\left\lceil \frac{n}{2}\right\rceil \in \mathbb{N}\, .
\end{equation}
Further clarifications on the suppression of the multi-Higgs amplitudes in SMEFT are provided in appendix~\ref{app:SMEFT-multiH-sup}. 
 
Amplitudes with a larger number of Higgs bosons in the final state require a higher number of internal propagators $N_I$. 
For this reason we can conclude that, in general, \textit{the production of multi-Higgs states is strongly suppressed in SMEFT}, unlike in generic HEFT theories. The first non-trivial contributions to each process scale as:
\begin{eqnarray} 
T_{\omega\omega\to 2h}\sim\frac{1}{\Lambda^2}\, ,\qquad  
T_{\omega\omega\to 3h, \, 4h}\sim\frac{1}{\Lambda^4}\, ,\qquad  
T_{\omega\omega\to 5h, \, 6h}\sim\frac{1}{\Lambda^6}\, ,\qquad \mbox{etc.} 
\label{eq:SMEFT-amp-suppression}
\end{eqnarray}
Note that higher SMEFT operators, with dimension $D>6$, are never providing an amplitude that dominates over the diagrams coming from the leading SMEFT Lagrangian~(\ref{eq:LSMEFT-dominant}). Thus, the suppression~(\ref{eq:SMEFT-amp-suppression}) is a general feature of SMEFT. 

This result fully agrees with our findings in the framework of the non-linear HEFT realization if one imposes the SMEFT correlations in the flare function $\mF(h)$~\cite{Gomez-Ambrosio:2022qsi}. A consistent SMEFT approach tells us that, e.g., the $1/\Lambda^4$ amplitudes are incomplete unless we incorporate the $\mL_{(D=8)}^{\rm SMEFT}$ interactions to the analysis, as they also produce $\mO(1/\Lambda^4)$ contributions to the scattering amplitude, as we have shown in previous subsections.

In contrast, \textit{non-SMEFT scenarios that allow the flare-function to deviate from those constraints can avoid this strong suppression in the multi-Higgs production cross sections}. 
We will illustrate this with some numerical simulations in the next sections.

\section{Cross section phenomenology} 
\label{sec:pheno}

In this section, we perform a phenomenological analysis of the previous results. We study the multi-Higgs cross sections in a scenario where the SMEFT provides a proper low-energy description and compare it with cases where it is not applicable, and only a HEFT description is appropriate. 
We will find an interesting suppression in SMEFT-compatible scenarios.

In general, we will show that the SMEFT cross sections are much smaller than those in the pure HEFT cases. This characteristic does not necessarily result from a large deviation from the SM couplings. Instead, we find that even slight deviations in the $a_j$'s correlations, typical of the SMEFT, can lead to a significant enhancement of the cross section, even when the deviations from the SM couplings are small.

For this study we will consider a series of models and benchmark points (BP). In particular, for SMEFT we will be using,
\begin{equation}
d\,=\, \frac{ 2 v^2 c_{H\Box}^{(6)} }{\Lambda^2}  \,=\, 0.1\,, 
\qquad \rho\, =\,  \, \ \frac{c_{H\Box}^{(8)}}{2 (c_{H\Box}^{(6)})^2} =\, 1\, .    
\end{equation}
Here, $d$ 
parametrizes the $D=6$ correction and it is related to the $hWW$ coupling $a=1+d/2$ at tree level. The chosen value $d=0.1$ is then related to the SM deviation $\Delta a=a-1=0.05$, within the range of the most precise experimental determinations up to date from ATLAS~\cite{ATLAS:2022vkf}, $a=\kappa_V=1.035\pm 0.031$, and CMS~\cite{CMS:2022dwd,hepdata.127765}, $a=\kappa_V=1.014\pm 0.029$.

The ratio $\rho=c_{H\Box}^{(8)}/[2 (c_{H\Box}^{(6)})^2]$ parametrizes the relation between $D=8$ and $D=6$ SMEFT Wilson coefficients, which has been taken as $\rho=1$ for all our BP. We will see that the precise values of these SMEFT parameters are not particularly relevant for the features observed in our HEFT-vs-SMEFT comparison.

We remark that, although the precise value of $\rho$ is not really relevant for the qualitative behaviour of our results (as far as $\rho\sim \mO(1)$), the choice of $d$ is crucial for the rate of convergence of the SMEFT power series. We could have chosen a BP with a much smaller $d$ (e.g., $d=0.01$). However this choice would have made the following examples less illustrative: the large differences we show in this section between SMEFT and non-SMEFT models would be even larger by several orders of magnitude; the convergence of the SMEFT power series would improve and $D=8$ corrections would be much more suppressed w.r.t. the $D=6$ ones; and the scanning of the cross sections in terms of the $a_3$ and $a_4$ couplings we later perform would be more difficult to visualize.    

\begin{figure}[t!] \label{CMS_region}   
     \centering
     \subfigure[]
 {
  \includegraphics[width=0.45\textwidth]{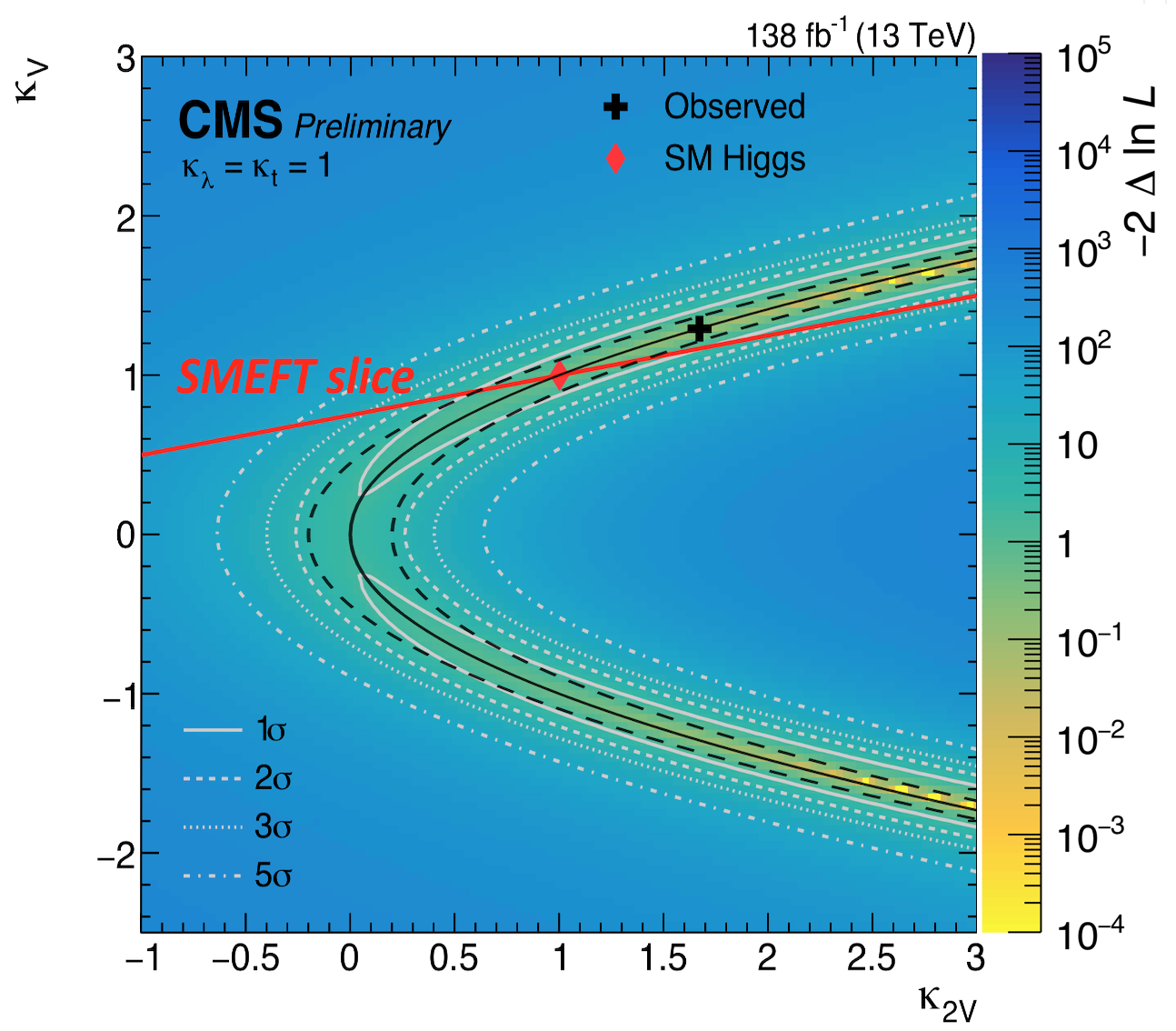} 
  }
 \subfigure[]
 {
  \includegraphics[width=0.49\textwidth]{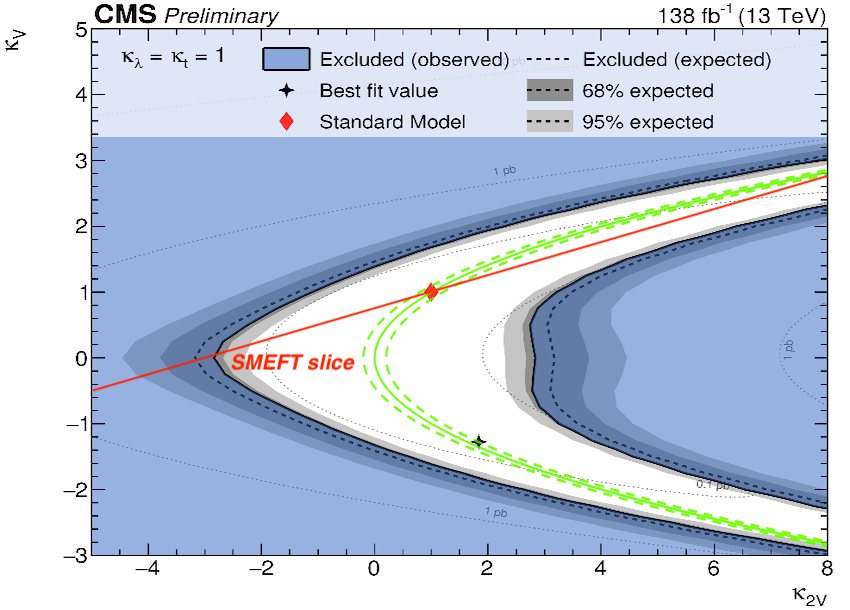} 
   }
\\ 
\vspace*{0.85cm}
  \subfigure[]
 {
  \includegraphics[width=0.45\textwidth]{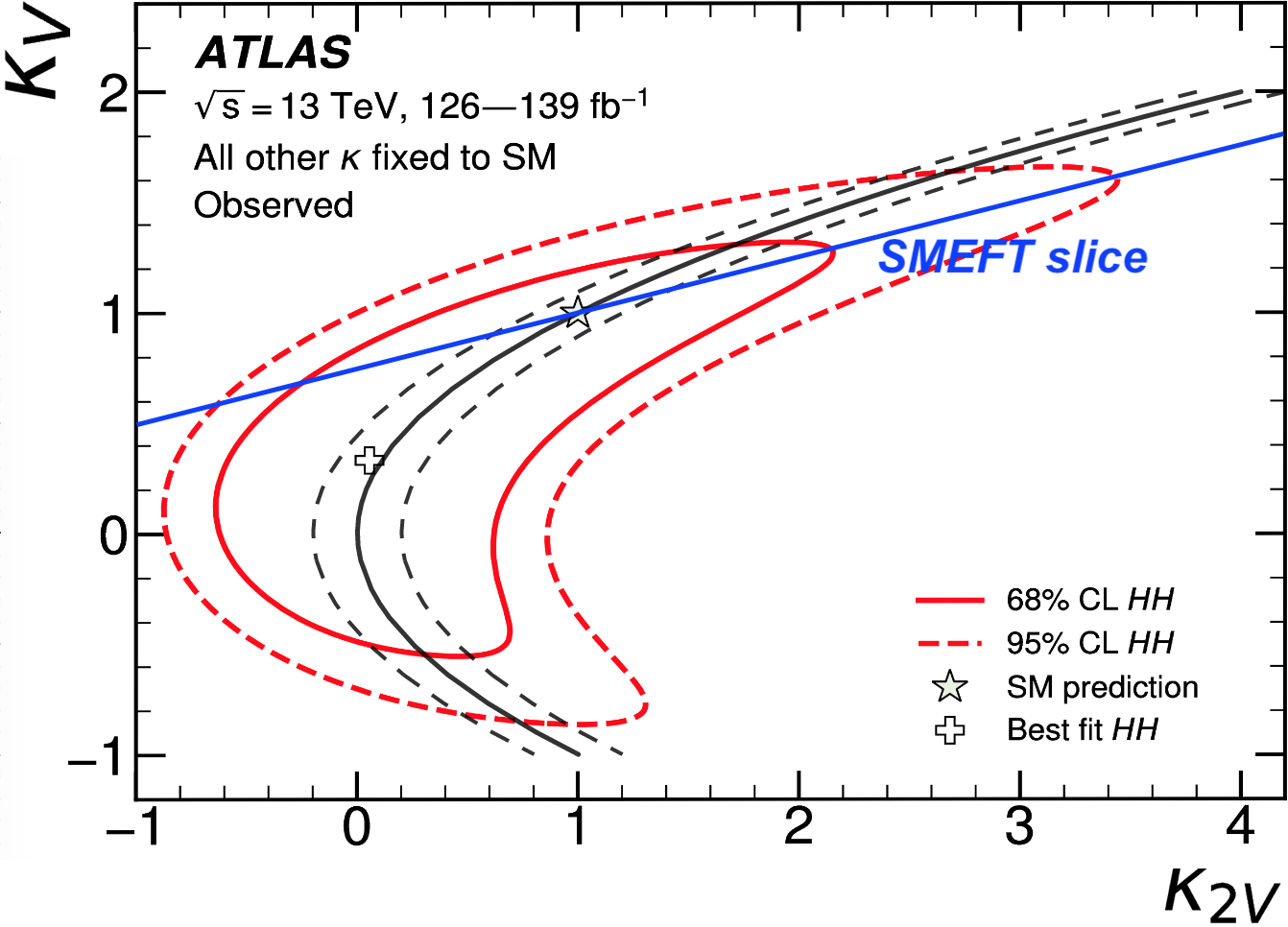}
  }
   \subfigure[]
  {
\includegraphics[width=0.49
   \textwidth]{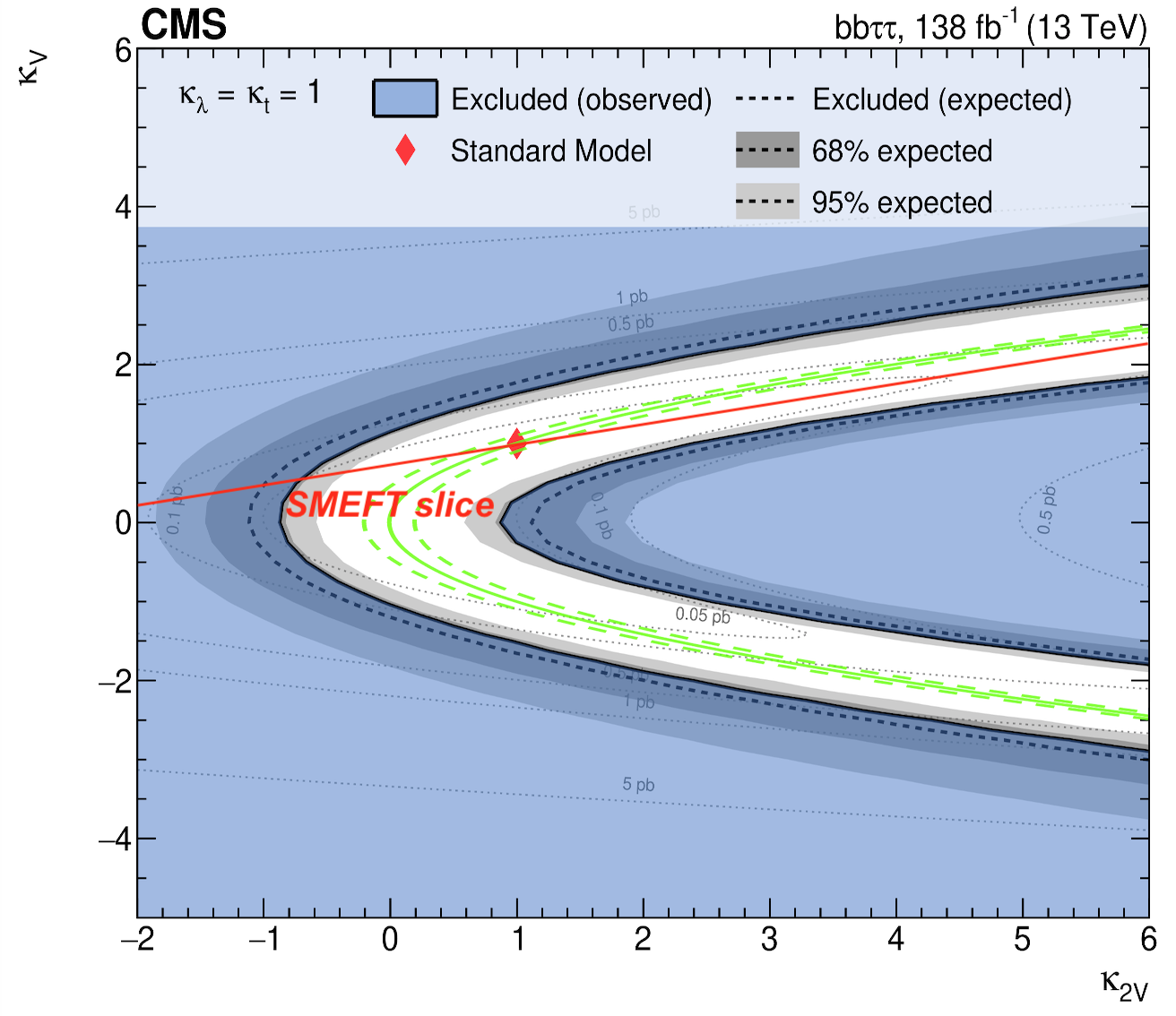}
   }
    \caption{{\small 
         {\bf a)}  CMS  
    experimental confidence regions for the $hWW$ coupling $\kappa_V=a=a_1/2$ and for the $hhWW$ coupling $\kappa_{2V}=b=a_2$, 
    from a non-resonant $hh$ production search with each Higgs boson decaying into a highly boosted $b\bar{b}$ pair~\cite{CMS:2022gjd}
    (white lines and colour map in figure~11 from the additional material in \href{https://cms-results.web.cern.ch/cms-results/public-results/publications/B2G-22-003/}{CMS-B2G-22-003}). 
    {\bf b)} 
    CMS confidence regions from processes with one Higgs boson decaying to $b\bar{b}$ and the other one to $W^+W^-$~\cite{CMS:2023qiw} (black lines and colour map).
    {\bf c)} 
    ATLAS experimental confidence regions,  
    from the combination of $hh\to b\bar{b}b\bar{b},\, b\bar{b}\tau^+\tau^-,\, b\bar{b}\gamma\gamma$ decay channels~\cite{ATLAS:2022kbf} (red lines).  
    {\bf d)} CMS searches for non-resonant production of Higgs boson pairs in $b\bar{b}\tau^+\tau^-$ final states~\cite{CMS:2022hgz} (black lines and colour map confidence regions).  
    We have superimposed the correlation $a_2= 2 a_1 - 3$  
 in SMEFT at $\mathcal{O}(1/\Lambda^2)$ (red line in {\bf a)}, {\bf b)} and {\bf d)}; blue line in {\bf c)}). In addition, we have also plotted the parabolas $\hat{a}_2=a_2-a_1^2/4 =0$~(solid black in {\bf a)} and {\bf c)}; solid green in {\bf b)} and {\bf d)}) and  
     $\hat{a}_2=\pm 0.2$~(dashed black in {\bf a)} and {\bf c)}; dashed green in {\bf b)} and {\bf d)}), which determine the longitudinal $WW\to hh$ scattering in the naive equivalence theorem.
    Reproduced under the Creative Commons BY 4.0 license.  
     }
    }
    \label{fig:CMS1}    
\end{figure}

The rationale behind the analysis presented here is as follows: we will consider benchmark points for different models and values of couplings which are very close to the SM and SMEFT in their first flare-function coefficients. 
 The coupling $a_1=2a$ is relatively well known from Higgs decay studies at the LHC. It is found to be close to the SM, up to $\mO(10\%)$ uncertainties.   
Current experimental determinations of $a_2=b$ from $hh$ production at the LHC~\cite{CMS:2023qiw,CMS:2022hgz,CMS:2020tkr,ATLAS:2020jgy,CMS:2022cpr,ATLAS:2022kbf,ATLAS:2023qzf,CMS:2022gjd,CMS:2022dwd} yield a very wide marginalized allowed range $a_2 \sim [\, 0\, , \, 2] $ (see appendix~\ref{app:a2-exp} for a detailed compilation of $a_2$ measurements).   
However, the combination of these results with the determinations for $a_1$ from Higgs decays will potentially reduce the uncertainty on $a_2$ to the $\mO(10\%)$ level.  
This can be observed in figure~\ref{fig:CMS1}, where the $hh$ production data is showing a fair enough determination of $\hat{a}_2=a_2-a_1^2/4$, with $|\hat{a}_2|\sim \mO(10\%)$ in the best cases~\cite{CMS:2022gjd} (this feature is easy to understand with the structure of our amplitudes and field redefinitions). Thus, even if the marginalized determinations of $a_2$ in those works 
carry a large dispersion, this is no longer true if one superimposes the experimental information on $a_1$.  

Recent phenomenological LHC studies~\cite{Anisha:2022ctm} have shown how these \emph{banana} shapes in the $(a_1,a_2)$-plane are expected to generally appear in future analyses, proving that we should eventually obtain a moderately good knowledge on $\hat{a}_2$. Thus, potentially stringent experimental determinations of $a_1$ and $\hat{a}_2$ can serve to discern and discard new physics models, even if the marginal uncertainty on $a_2$ remains sizeable in the near future.      
For this reason, we will consider two types of BPs: one with values of $a_1$ within the experimentally allowed range while keeping $a_2$ free, and a second type in which both $a_1$ and $a_2$ are constrained to have small deviations from the Standard Model.


In the literature one can find several works that discuss specific UV-completions for the Higgs sector. 
Some of them accept a SMEFT description and some don't.
However, to distinguish between them, it is not enough to know whether the model shows a non-linear HEFT structure, where the Higgs boson is a Goldstone from some BSM symmetry, or whether the EFT derives from some strongly-interacting underlying theory.

A paradigmatic example is that of the minimally composite Higgs Model~\cite{Agashe:2004rs,Contino:2011np}, in which the EW Goldstones and the Higgs boson have a composite structure and evident non-linear transformation properties, and nontheless still be described by a SMEFT-like Lagrangian for a compositeness scale $f\gg v$~\cite{Alonso:2015fsp,Alonso:2016btr,Alonso:2016oah}.

Another example is that of the dilaton model~\cite{Halyo:1991pc,Goldberger:2007zk,Vecchi:2010gj,Chacko:2012sy,Bellazzini:2012vz}, where the Higgs is a (pseudo) Goldstone from the spontaneous conformal symmetry breaking. This model does not have a SMEFT description, since the theory does not obey the required flare function regularity constraints~\cite{Gomez-Ambrosio:2022qsi,Salas-Bernardez:2020hua}. Still, in this model, the $\omega\omega\to n \times h$ cross sections vanish in the EqTh approximation for any number of final Higgs particles (see appendix~\ref{app:field-redef} for details), and becomes of little interest for our purposes. 

Alternatively, there are UV-complete models with perturbative coupling expansions which give place to pure-HEFT low-energy theories, and cannot be described by SMEFT. Singlet-scalar extensions of the SM can lead to a pure-HEFT picture if the singlet's mass and its vacuum expectation value get a large enough separation~\cite{Buchalla:2016bse,Boggia:2016asg,Cohen:2020xca}. Also, the 2-Higgs Doublet Model can produce a non-SMEFT low-energy theory for certain choices of its parameters~\cite{Arco:2023sac}. These studies ensure the preservation of unitarity and perturbativity; nonetheless, the integration of the heavy degrees of freedom results in an EFT expansion that cannot be straightforwardly organized in terms of the canonical dimension of the operators.

The structure of these particular models is indeed quite involved and an exhaustive analysis of the various non-SMEFT cases therein is far beyond the scope of the present work.   
In this article, we will illustrate our phenomenological studies with the simplified flare function models considered in the next subsection.

\subsection{Models and benchmark points}

 Here we present the analytical expressions for each $a_j$ in terms of the specified numerical inputs for every BP. To illustrate, we provide approximate numerical values for these $a_j$ couplings, although their precise values can always be readily determined using the BP analytical relations. We will examine the following cases:

\begin{itemize}

\item{\bf SMEFT:} 
for this BP we will be using the previously selected inputs 
$ d\,=\, 0.1\,,$ $\rho\, =\, 1\,  $ in (\ref{eq:SMEFT-FF}).

\begin{itemize}

\item{\bf SMEFT$^{\rm (D=6)}$:} 
At dimension $D=6$ approximation, the considered BP, $ d\,=\, 0.1\,,$ yields the values for $\omega\omega\to n \times h$ vertex couplings: 
\begin{equation}
a = \frac{a_1}{2} 
=1.05\, ,\quad 
b= a_2  
=1.20\,,\quad 
a_3  
= 0.1\wideparen{3}\,,\quad
a_4  
= 0.0\wideparen{3}\, , 
\label{eq:BP-SMEFT6}
\end{equation}
with $a_{k\geq 5}=0$ for all the remaining couplings.

\item[-]{\bf SMEFT$^{\rm (D=8)}$:} 
At dimension $D=8$ approximation, the considered BP, $ d\,=\, 0.1\,,$ and $\rho=1$, provides for the $\omega\omega\to n \times h$ vertex couplings the values: 
\begin{equation}
a = \frac{a_1}{2} 
\approx 1.06 
\, , \, \,\, 
b= a_2  
=1.26\,, \, \,\, 
a_3 
=0.22\,, \, \,\,
a_4  
=0.10\, ,
\label{eq:BP-SMEFT8}
\end{equation}
with the couplings $a_5  
=0.02\wideparen{6}$, $a_6  
=0.00\wideparen{4}$ and $a_{k\geq 7}=0$, irrelevant for this article.

\end{itemize}

\item{\bf Non-SMEFT BP1: exponential flare function $\mF(h)=\exp{ \bigg\{ f(h) \bigg\} }$. }  

In this scenario, we provide a flare function, $\mathcal{F}(h)$, with no real zeroes (given that $f(h)$ is a polynomial). It describes models which do not accept a SMEFT description~\cite{Alonso:2016oah,Cohen:2020xca,Gomez-Ambrosio:2022qsi,Gomez-Ambrosio:2022why}, but fulfil the positivity requirements pointed out in \cite{Gomez-Ambrosio:2022qsi}. Furthermore, the general term $a_n$ in $\mF(h)$ is going to have a behaviour of the form $a_n\sim \frac{1}{n!}$, decreasing and vanishing at large $n$. 

For a fair comparison with the SMEFT BP, we will match the first coefficients to those in the $D=6$ expressions in~(\ref{eq:BP-SMEFT6}). Nevertheless, we will see that it is not really crucial for the results to consider the $D=6$ approximation as similar results are obtained if one considers either the $D=8$ expressions~(\ref{eq:BP-SMEFT8}) or the SM values $a_2=a_1/2=1$, $a_{k\geq 3}=0$.

\begin{itemize}
\item[-]{\bf BP1$^{\rm (a_1)}$:} 
We consider the simplest exponential flare function that matches the $D=6$ SMEFT prediction in~(\ref{eq:BP-SMEFT6}) for $a_1$:  
\begin{eqnarray}
&&\mF(h)=\exp{\left\{ a_1 \frac{h}{v}\right\}  } 
\qquad \Longrightarrow \qquad 
\label{eq:BP1-a1}
\\
&& \qquad  
b= a_2 = \frac{a_1^2}{2!} 
=2.205 \,,\,\,\,  
a_3= \frac{a_1^3}{3!} 
\approx 1.54 \,,\,\,\, 
a_4 = \frac{a_1^4}{4!}  
\approx 0.81  
\, ,
\nonumber
\end{eqnarray}
where we are using the input $a  = \frac{a_1}{2} 
=1.05$ from~(\ref{eq:BP-SMEFT6}).

\item[-]{\bf BP1$^{\rm (a_1,\, a_2)}$:} 
In a second approach we consider the simplest exponential flare function that matches the $D=6$ SMEFT prediction in~(\ref{eq:BP-SMEFT6}) for $a_1$ and $a_2$:    
\begin{eqnarray}
&&\mF(h)=\exp{\left\{ a_1 \frac{h}{v} \, +\, \left(a_2-\frac{a_1^2}{2}\right)\frac{h^2}{v^2}  \right\}  } 
\qquad \Longrightarrow \qquad 
\label{eq:BP1-a1-a2}
\\
&&   
 a_3=\left(a_1 a_2 -\frac{a_1^3}{3}\right) 
\approx -0.57 \,,\qquad
a_4 = \left(\frac{a_2^2}{2} -\frac{a_1^4}{12}\right) 
\approx -0.90\, ,  
\nonumber
\end{eqnarray} 
where we are using the input $a  = \frac{a_1}{2}  
 =1.05$ and 
$b= a_2  
=1.20$ from~(\ref{eq:BP-SMEFT6}).

\end{itemize}

\item{\bf Non-SMEFT BP2: rational flare function $\mF(h)=\bigg(1\, -\,   f(h) \bigg)^{-2}$. }  

In this scenario we provide, in terms of a polynomial $f(h)$, a flare function which also has no real zeroes and, hence, the corresponding model does not accept a SMEFT description. By construction, this type of models also fulfill the positivity requirements~\cite{Gomez-Ambrosio:2022qsi}. 
In this case, however, contrary to the exponential case,  we will be dealing with general terms $a_n$ in $\mF(h)$ that turn large as $n$ grows, with a behaviour $a_n\sim n$. 
Again, for an appropriate comparison with the SMEFT BP, we will match the first coefficients to those in the $D=6$ expressions in~(\ref{eq:BP-SMEFT6}). As it happened with the BP1 models, we will see that the precise value of $a_1$ and $a_2$ is not critical for the conclusions drawn here, as long as these couplings are close to their SM limit.

\begin{itemize}
\item[-]{\bf BP2$^{\rm (a_1)}$:} 
We consider the simplest rational flare function that coincides with the $D=6$ SMEFT prediction in~(\ref{eq:BP-SMEFT6}) for the value of $a_1$:    
\begin{eqnarray}
&&\mF(h)= \left(1\,-\,  \frac{a_1}{2}  \frac{h}{v}   \right)^{-2}  
\qquad \Longrightarrow \qquad 
\label{eq:BP2-a1}
\\
&& 
    b= \frac{3}{4}a_1^2  
\approx 3.31\,,\qquad 
a_3= \frac{1}{2}a_1^3  
\approx 4.63\,,\qquad    
a_4 = \frac{5}{16}a_1^4    
\approx 6.08\, ,    
\nonumber
\end{eqnarray} 
where we are using the input $a  = \frac{a_1}{2} 
=1.05$ from~(\ref{eq:BP-SMEFT6}).

\item[-]{\bf BP2$^{\rm (a_1,\, a_2)}$:} 
In a second approach we consider the simplest rational flare function that matches the $D=6$ SMEFT prediction in~(\ref{eq:BP-SMEFT6}) for the values of $a_1$ and $a_2$:    
\begin{eqnarray}
&&\mF(h)= \left(1\,-\,  \frac{a_1}{2} \frac{h}{v} \, -\, \left(\frac{a_2}{2}-\frac{3a_1^2}{8}\right)\frac{h^2}{v^2}  \right)^{-2}  
\qquad \Longrightarrow \qquad 
\label{eq:BP2-a1-a2}
\\
&& 
  a_3= \frac{1}{8}\left(-5 a_1^3+12 a_1 a_2\right)  
\approx -2.01\,,\qquad  
\nonumber\\
&&
 a_4 =\frac{1}{64}\left(-25 a_1^4+ 24 a_1^2 a_2 + 48 a_2^2\right) 
\approx -4.53\, ,    
\nonumber
\end{eqnarray} 
where we are using the input $a  = \frac{a_1}{2} =1.05$ and 
$b= a_2  
=1.20$ from~(\ref{eq:BP-SMEFT6}).

\end{itemize}

\end{itemize}

\subsection{BP study for \texorpdfstring{$\omega\omega\to 2h$}{ww->2h}}

\begin{figure}[t!] 
\center
  \includegraphics[width=0.8\textwidth]{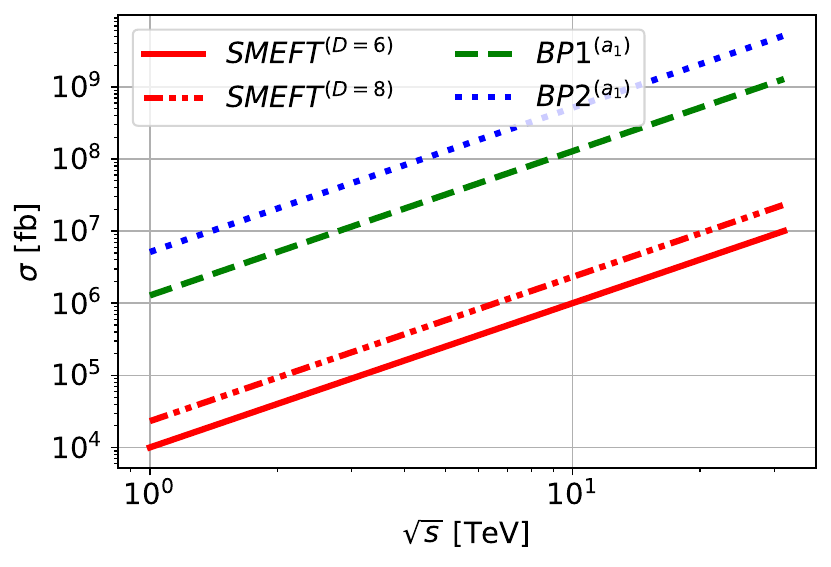}
    \caption{{\small 
    Comparison of the $\omega\omega\to 2h$ cross section predictions of SMEFT$^{\rm (D=6)}$,  BP1$^{\rm (a_1)}$ and BP2$^{\rm (a_1)}$. For illustration, we also provide the results for SMEFT$^{\rm (D=8)}$, which refines the previous order SMEFT determination. The comparison of non-SMEFT BP's with both SMEFT$^{\rm (D=6)}$ and SMEFT$^{\rm (D=8)}$ casts similar conclusions: SMEFT cross sections are suppressed with respect to non-SMEFT models.      
    }}
    \label{fig:pheno-2h}     
\end{figure}

Figure~\ref{fig:pheno-2h} shows the expected $\omega\omega\to 2h$ cross sections for SMEFT. We are using the $D=6$ coupling values~(\ref{eq:BP-SMEFT6}). Dimension $D=8$ corrections are further suppressed and lead to a similar outcome. We show the $(D=6)$-vs-$(D=8)$ comparison in this plot for illustration but we will no further discuss the SMEFT$^{\rm (D=8)}$ in what follows, only the $D=6$ predictions.

SMEFT is then compared in figure~\ref{fig:pheno-2h} with the two non-SMEFT BP's discussed in this article. We plot the predictions for BP1$^{\rm (a_1)}$ and BP2$^{\rm (a_1)}$.  
The refined BP's,  BP1$^{\rm (a_1,a_2)}$ and BP2$^{\rm (a_1,a_2)}$, are not discussed in the plot. Since by construction $a_1$ and $a_2$ are set to the SMEFT$^{\rm (D=6)}$ values in~(\ref{eq:BP-SMEFT6}), they yield the same $\omega\omega\to 2h$ cross section as SMEFT$^{\rm (D=6)}$.

Our conclusion is that, as expected from previous sections, the SMEFT cross section is suppressed by several orders of magnitude with respect to the non-SMEFT ones: in general, we will see that the fractional flare functions (BP2) tend to yield higher cross section than the exponential ones (BP1), both of them being well over the SMEFT determinations.

It is important to remark that the precise input $a_1=2a=2.1$ is not critical for the non-SMEFT models BP1 and BP2, obtaining essentially the same cross sections if one considers the SM limit $a_1=2 a=2$. This is not true for SMEFT, which relies on very fine-tuned relations between the $a_j$ effective couplings~\cite{Gomez-Ambrosio:2022qsi,Gomez-Ambrosio:2022why}. In particular, imposing the SM $a_1$ value in SMEFT$^{\rm (D=6)}$ leads to $\hat{a}_2=0$ and a total suppression of the cross section.

\subsection{BP study for \texorpdfstring{$\omega\omega \to 3h$}{ww->3h}}

\begin{figure}[t!]   
\center
  \includegraphics[width=0.8\textwidth]{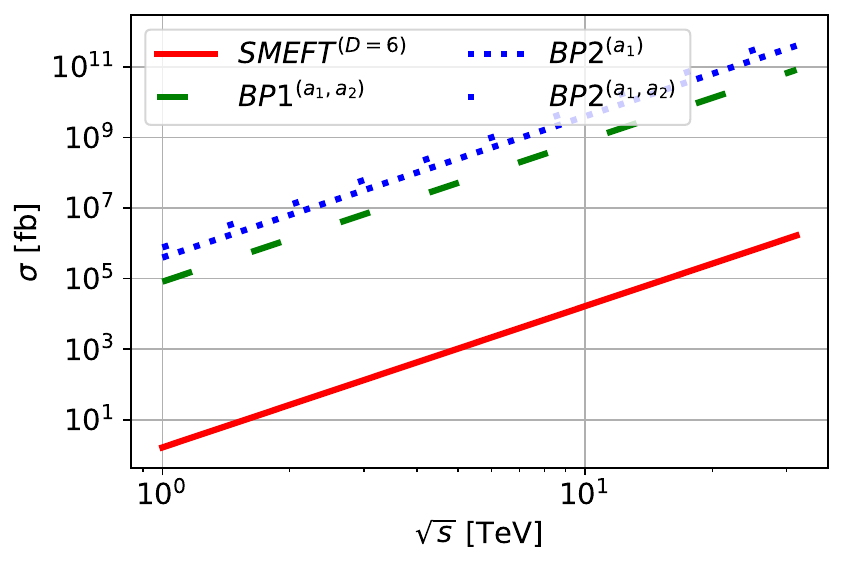}
    \caption{{\small  Comparison of the $\omega\omega\to 3h$ cross section predictions of SMEFT$^{\rm (D=6)}$,   
    BP1$^{\rm (a_1,a_2)}$, BP2$^{\rm (a_1)}$ and  BP2$^{\rm (a_1,a_2)}$. We note that the $\omega\omega\to 3h$ cross section is identically zero for the first exponential model, BP1$^{\rm (a_1)}$, regardless of the value considered for the $a_1$ input and, therefore, it has not been plotted.}}
    \label{fig:pheno-3h}     
\end{figure}

Figure~\ref{fig:pheno-3h} shows our predictions for the  $\omega\omega\to 3h$ cross sections for SMEFT, where we are using the $D=6$ coupling values given in~(\ref{eq:BP-SMEFT6}). This prediction is then compared in that plot with the pure-HEFT scenarios, the two non-SMEFT BP's discussed in this article. We plot the predictions for BP1$^{\rm (a_1)}$ and BP2$^{\rm (a_1)}$.  

In this case, we are also showing the predictions for the refined BP's, BP1$^{\rm (a_1,a_2)}$ and BP2$^{\rm (a_1,a_2)}$, as the $\omega\omega\to 3h$ cross section now also depends on $a_3$. 
In the case of the BP2 fractional flare functions we find a very similar result and a cross section 5 orders of magnitude larger than the SMEFT one. 

 The BP1 exponential flare functions show an interesting peculiar behaviour. The BP1$^{\rm (a_1)}$ model, with $\mF(h)=\exp\left\{ a_1 h/v\right\}$ leads to a $\omega\omega\to 3h$ amplitude that vanishes for any value of $a_1$. Thus, it has not been represented in the logarithmic plot of figure~\ref{fig:pheno-3h}. On the other hand, the cross sections for other ($n\neq 3$) multiplicities lead to a large cross section with respect to the SMEFT (see Figs.~\ref{fig:pheno-2h} and~\ref{fig:pheno-4h}). Although this may resemble the dilaton-Higgs case ($a_2=b\,=\, a=a_1/2$ and $a_{k\geq 3}=0$)~\cite{Halyo:1991pc,Goldberger:2007zk,Vecchi:2010gj,Chacko:2012sy,Bellazzini:2012vz}, such a model leads to a vanishing $\omega\omega\to n\times  h$ cross section for any number $n$ of final Higgs bosons. 
This pathology is cured in the refined exponential model BP1$^{\rm (a_1,a_2)}$, which now provides again a sizable cross section, orders of magnitude larger than the SMEFT one.

The former discussion shows indeed the full generality of the HEFT approach. This implies  that, in most cases, the presence of `arbitrary' $\mO(1)$ $a_j$ couplings leads to large cross sections in comparison with those obtained in SMEFT calculations. In the latter, the $a_1$ and $a_2$ couplings are $\mO(1)$ but there are very fine-tuned correlations among all the $a_j$'s~\cite{Gomez-Ambrosio:2022qsi,Gomez-Ambrosio:2022why}, leading to a very suppressed SMEFT cross section for multi-Higgs production.

An interesting fact of the HEFT models is that, although they typically  lead to large cross sections, this is not always necessarily true.  There are certain cases with cross sections even more suppressed than the SMEFT one, for instance, the dilaton model.

Our conclusions are that, as expected, the SMEFT cross section is suppressed by several orders of magnitude with respect to the non-SMEFT ones: in general, we will see that the fractional flare functions (BP2) tend to yield higher cross sections than the exponential ones (BP1), being both of them well over the SMEFT determinations.

It is important to remark that the precise input $a_1=2a=2.1$ is not critical for the non-SMEFT models BP1 and BP2, obtaining essentially the same cross sections if one considers the SM limit $a_1=2 a=2$. This is not true for SMEFT, which relies on very precise relations between the $a_j$ effective couplings~\cite{Gomez-Ambrosio:2022qsi,Gomez-Ambrosio:2022why}. In particular, imposing the SM $a_1$ value in SMEFT$^{\rm (D=6)}$ leads to $\hat{a}_2=0$ and a total suppression of the cross section. 

\begin{figure}[ht!]
\center
  \includegraphics[width=0.8\textwidth]{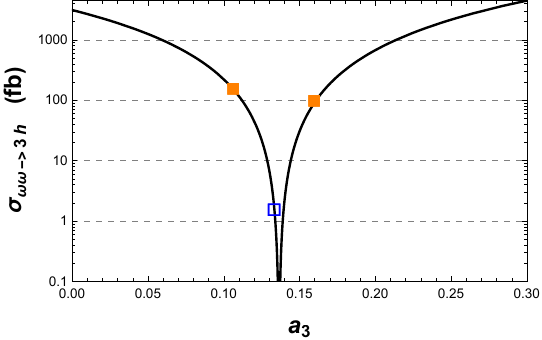}
    \caption{{\small Scan of the $\omega\omega\to 3h$ cross section predictions in terms of $a_3$ at $\sqrt{s}=1$~TeV. The inputs $a_1=a_1^{\rm SMEFT (D=6)}=2.1$ and $a_2=a_2^{\rm SMEFT (D=6)}=1.2$ are taken from~(\ref{eq:BP-SMEFT6}), the SMEFT$^{\rm (D=6)}$ BP. We have marked a few especial points: $a_3=a_3^{\rm SMEFT (D=6)}=0.1\wideparen{3}$ (empty blue square) and their 20\% deviations (full orange squares), $a_3=80\% \times a_3^{\rm SMEFT (D=6)}$ and  $a_3=120\% \times a_3^{\rm SMEFT (D=6)}$. We note that, in between, $\sigma_{\omega\omega\to 3h}$ vanishes at $a_3=\frac{2}{3} a_1 \left(a_2-\frac{1}{4}a_1^2\right)= 0.1365$.  
    }}
    \label{fig:scan-pheno-3h}     
\end{figure}

To end this subsection, figure~\ref{fig:scan-pheno-3h} presents a scan of the $\omega\omega \to 3h$ cross-section in terms of $a_3$, analogous to that performed in ref.~\cite{Englert:2023uug} for $\omega\omega \to 2h$ with $a_2$. For the numerical analysis we set $a_1$ and $a_2$ to the SMEFT$^{\rm (D=6)}$ values in~(\ref{eq:BP-SMEFT6}).  One can observe that the cross section has a zero for $a_3=\frac{2}{3} a_1 \left(a_2-\frac{1}{4}a_1^2\right)\approx 0.14$ (for these numerical inputs). Close by, one finds the SMEFT$^{\rm (D=6)}$ BP. But one can easily observe how fine-tuned this value is: a $\pm20\%$ variation in $a_3$ w.r.t. the  SMEFT$^{\rm (D=6)}$ value leads to a rising in the cross section of orders of magnitude, even if $a_1$ and $a_2$ are left unchanged. Obviously, $\mO(1)$ values for $a_3$ (such as, e.g., those in BP1$^{\rm (a_1,a_2)}$ in~(\ref{eq:BP1-a1-a2}) and  BP2$^{\rm (a_1,a_2)}$ in~(\ref{eq:BP2-a1-a2}) produce even larger cross sections, as one can see in figure~\ref{fig:pheno-3h}.

\subsection{\texorpdfstring{BP study to $\omega\omega \to 4h$}{ww->4h}}

\begin{figure}[t!]  
\center
  \includegraphics[width=0.8\textwidth]{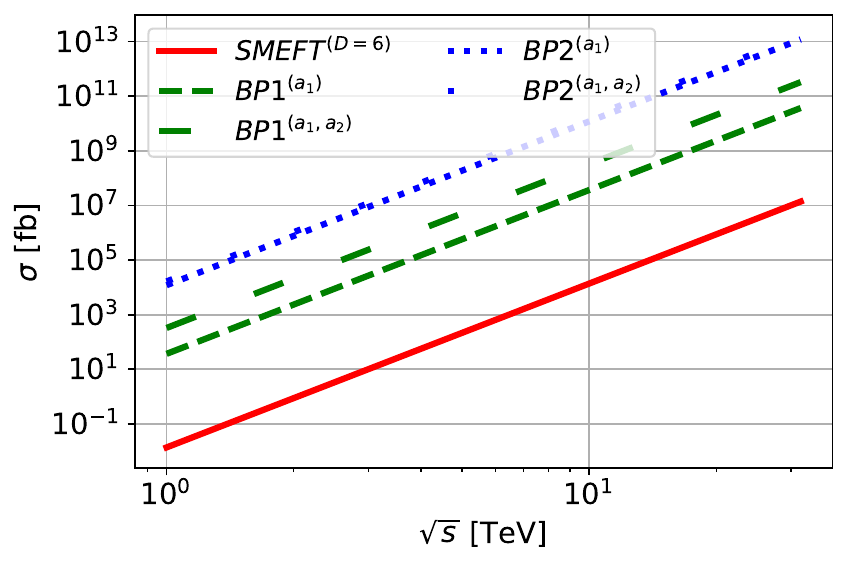}
    \caption{{\small  Comparison of the $\omega\omega\to 4h$ cross section predictions of SMEFT$^{\rm (D=6)}$,  
    BP1$^{\rm (a_1)}$,  BP1$^{\rm (a_1,a_2)}$, BP2$^{\rm (a_1)}$ and  BP2$^{\rm (a_1,a_2)}$. 
    }}
    \label{fig:pheno-4h}     
\end{figure}

Figure~\ref{fig:pheno-4h} shows the comparison between the SMEFT$^{\rm (D=6)}$ BP and the non-SMEFT BP's, BP1 and BP2. One can observe that in this case the results from matching just the $a_1$ coupling in~(\ref{eq:BP-SMEFT6}) (BP1$^{(a_1)}$ and BP2$^{(a_1)}$) or matching $a_1$ and $a_2$ (BP1$^{(a_1,a_2)}$ and BP2$^{(a_1,a_2)}$) are close to each other, respectively. It is also important to notice that on the contrary to want happens with the SMEFT BP, which works on very fine-tuned cancellations, the results for the non-SMEFT BP's are very stable: one could set the SM values $a_1=2$ and $a_2=1$ and the predictions for BP1 and BP2 remain essentially unchanged.

\begin{figure}[t!] 
\center
  \includegraphics[width=0.8\textwidth]{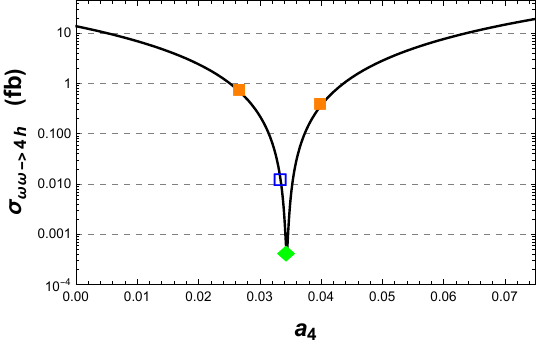}
    \caption{{\small Scanning of the $\omega\omega\to 4h$ cross section predictions in terms of $a_4$ at $\sqrt{s}=1$~TeV. The inputs $a_1=a_1^{\rm SMEFT (D=6)}=2.1$, $a_2=a_2^{\rm SMEFT (D=6)}=1.2$ and  $a_3=a_3^{\rm SMEFT (D=6)}=0.1\wideparen{3}$ are taken from~(\ref{eq:BP-SMEFT6}), the SMEFT$^{\rm (D=6)}$ BP. We have marked a few especial points: $a_4=a_4^{\rm SMEFT (D=6)}=0.0\wideparen{3}$ (empty blue square) and their 20\% deviations (full orange squares), $a_4=80\% \times a_4^{\rm SMEFT (D=6)}$ and  $a_4=120\% \times a_4^{\rm SMEFT (D=6)}$. 
     The cross section's minimum is not zero this time and it is found at $a_4= \frac{3}{4}a_1a_3 - \frac{5}{12}a_1^2 \hat{a}_2 +\frac{1}{3}\hat{a}_2^2 (1-\chi_1)\approx 0.0344$ (filled green diamond).}
    }
    \label{fig:scan-pheno-4h}     
\end{figure}

In general one can observe that, as expected, the non-SMEFT cross sections are orders of magnitude larger than the SMEFT one. As mentioned above, the $a_j$ couplings are highly correlated in SMEFT, so the outcome is extremely sensitive to the precise value of each of them. Small variations on any of the relevant $a_j$ from the values prescribed by SMEFT lead to orders of magnitude modifications in the $\omega\omega \to 4h$ cross section. We illustrate this through the $a_4$ scan in figure~\ref{fig:scan-pheno-4h}, analogous to the $a_2$ scan for $\omega\omega\to hh$ in~\cite{Englert:2023uug}. If one considers for the SMEFT$^{\rm (D=6)}$ BP in~(\ref{eq:BP-SMEFT6}) ($a_1=2.1$, $a_2=1.2$, $a_3=0.1\wideparen{3}$, $a_4=0.0\wideparen{3}$) one obtains a very suppressed crossed section $\sigma_{\omega\omega \to 4h}\sim 10^{-2}$~fb. 
We will now leave $a_{1,2,3}$ fixed to the SMEFT$^{\rm (D=6)}$ values and vary $a_4$. 
A slight increase in $a_4$ from its SMEFT$^{\rm (D=6)}$ value leads to a decreasing of the cross section by an order of magnitude.    
However, in this case its minimum is not zero and it is found at $a_4= \frac{3}{4}a_1a_3 - \frac{5}{12}a_1^2 \hat{a}_2 +\frac{1}{3}\hat{a}_2^2 (1-\chi_1)\approx 0.0344$. From that point on, one can see that increasing or lowering $a_4$ leads to a rising of the cross section by several orders of magnitude. Figure~\ref{fig:scan-pheno-4h} shows that a $\pm 20\%$ variation in $a_4$ w.r.t. $a_4^{\rm SMEFT (D=6)}$ increases the cross section by two orders of magnitude.

\section{Exclusion plots for SMEFT scenarios}
\label{sec:SMEFT-exclusion} 


In this section we will assume that the underlying UV theory accepts a SMEFT description in the IR. We plan to provide some illustrative estimates of how large the $\omega\omega\to n\times h$ cross sections can be according to the present experimental observations for $a_1$~\cite{ATLAS:2022vkf,CMS:2022dwd} and $a_2$~\cite{ATLAS:2022kbf,CMS:2023qiw,CMS:2022gjd}. More precisely, current LHC $hh$ production analyses have actually shown an important sensitivity to $\hat{a}_2=a_2-a_1^2/4$, as one can see in figure~\ref{fig:CMS1} (see also~\cite{Anisha:2022ctm}).

This article is focused on the physics of the hard subprocesses $\omega\omega\to n\times h$ and a full simulation for LHC and future colliders is left for a future work. 
For this reason, we will just consider some illustrative rough bounds from $h\to WW$~\cite{ATLAS:2022vkf,CMS:2022dwd} and double-Higgs production at the LHC (figure~\ref{fig:CMS1}): 
\begin{equation} 
|\Delta a | \lsim 0.05 \, ,\qquad |\hat{a}_2|\lsim 0.1\, ,
\end{equation}
with $ \Delta a   =  a-1$ and $\hat{a}_2=b-a^2$.

In the study in this section we will consider these inequalities under the SMEFT perspective and use them to provide predictions for multi-Higgs production cross sections. For this, we will consider the contribution to the amplitudes from SMEFT operators at the corresponding lowest non-trivial order ($1/\Lambda^2$ for $2h$, $1/\Lambda^4$ for $3h$, etc.). Higher orders introduce corrections which might be relevant for high precision physics but are not going to teach us much here.    

In first place, under SMEFT, an experimental restriction on the $h\to WW$ coupling implies also a limitation on the allowed range for $a_2$ and, more specifically, on $\hat{a}_2$. At lowest order in the SMEFT expansion, this implies:  
\begin{eqnarray}
- 0.05  \leq \Delta a =  \frac{1}{2}d  \leq 0.05 
\qquad\Longrightarrow\qquad 
    -0.1 \leq \hat{a}_2  = d  \leq 0.1 \, . \label{eq:a2hat-bound}
\end{eqnarray} 
One can readily see that this implies a bound on the SMEFT $D=6$ Wilson coefficient, parametrized by $d\equiv 2v^2 c_{H\Box}^{(6)}/\Lambda^2$ in~(\ref{eq:d-rho-defs}). 

Nothing is known about the coupling $a_3$ (or $a_4$, $a_5$, etc.), although triple-Higgs production analyses such as~\cite{Gonzalez-Lopez:2020lpd} can be used to assess the sensitivity of current and future colliders to that parameter, $a_3$.   
Indeed, the relevant parameters for $3h$ and $4h$ production are not actually $a_3$ and $a_4$ but rather  
the combinations $\hat{a}_3= \frac{4}{3} d^2 (1+\rho)+ \mO(d^3)$ and $\hat{a}_4=\frac{1}{3} d^2  (1+\rho)+ \mO(d^3)$ provided in eq.~(\ref{eq:SMEFT-ajhat}). 
One can identify the two types of SMEFT contributions to  $\omega\omega\to 3h$ and $\omega\omega\to 4h$: a single insertion of one $D=8$ operator proportional to $c_{H\Box}^{(8)}$ parametrized by $\rho\equiv c_{H\Box}^{(8)}/\left[2(c_{H\Box}^{(6)})^2\right]$ ($d^2\rho$ terms); and a double insertion of $D=6$ operators proportional to $\left(c_{H\Box}^{(6)}\right)^2$ ($d^2$ terms without $\rho$).  
In the numerical analysis that follows we will be taking the assumption that the $ c_{H\Box}^{(8)}$  terms are similar in size to the $(c_{H\Box}^{(6)})^2$ contributions, or smaller. The case where  $ c_{H\Box}^{(8)}$ dominates over  $(c_{H\Box}^{(6)})^2$ --or the latter is absent~\cite{Dawson:2022cmu}-- leads to a very different phenomenology driven by the $1/\Lambda^4$ corrections in the $a_j$'s, where they now follow different correlations. This scenario is discussed in appendix~\ref{app:LO-SMEFT8} but will not be further analyzed here. 
For this reason, for the illustration in this section, we will consider for our numerical study the simple bounds:
\begin{equation}
   |d|\, \leq \, d_{\rm max}\, =\, 0.1\,, \qquad  |\rho|\, \leq \, \rho_{\rm max}\, =\, 1\, .
\label{eq:allowed-d-rho}
\end{equation}

We now consider the allowed values of the parameters $d$ and $\rho$ and maximize the SMEFT $2h$, $3h$ and $4h$ cross sections at their lowest non-trivial order in the EFT expansion:
\begin{eqnarray}  
\sigma_{\omega\omega\to 2h} &=&  \frac{8\pi^3}{s} \, d^2  \, \left(\frac{s}{16\pi^2 v^2}\right)^2 \, , 
\\
\sigma_{\omega\omega\to 3h}&=&  \frac{64\pi^3}{ 3s}   \, d^4\, (1+\rho)^2 \,  \left(\frac{s}{16\pi^2 v^2}\right)^3  \, , 
\\
\sigma_{\omega\omega\to 4h}&=&       \frac{8\pi^3}{9 s}\left(\frac{s}{16\pi^2 v^2}\right)^4  \, d^4\,  \bigg[  (1+\rho)^2 + 2(1+\rho) \chi_1  + \chi_2  
\bigg]\, .  
\end{eqnarray} 
For each energy the cross section maxima are reached for $d=d_{\rm max}$ and $\rho=\rho_{\rm max}$.  

\begin{figure}[t!] 
    \centering
 {
  \includegraphics[width=0.8\textwidth]{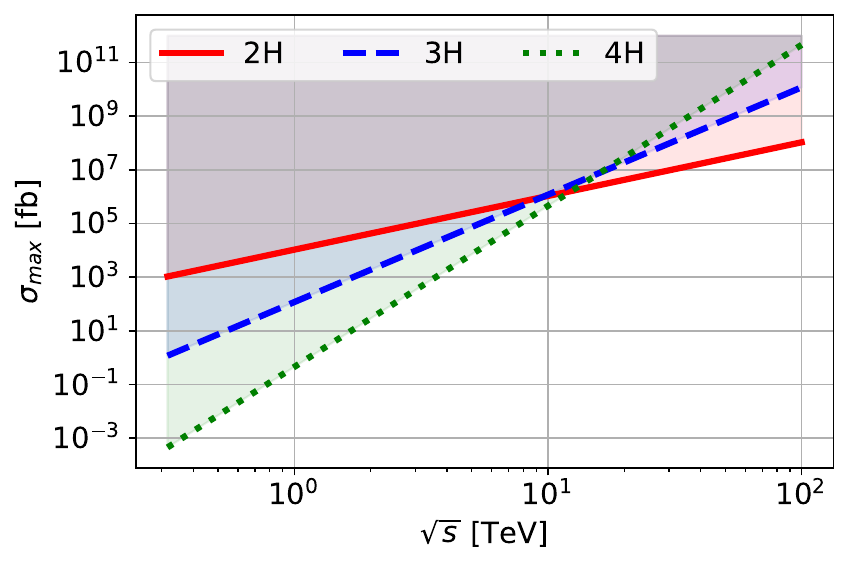}
  \vspace*{0.5cm}
  }
  \caption{SMEFT exclusion plot for the cross sections for 2, 3 and 4 Higgs bosons with $|d|\, \leq \, d_{\rm max}\, =\, 0.1$ and $|\rho|\, \leq \, \rho_{\rm max}\, =\, 1$.  
The regions above the solid, dashed and dotted lines can be safely excluded if the Wilson coefficients are within the considered range. 
Notice that the EFT perturbativity condition is not considered in this figure, as the EFT expansion breaks down on the region past the crossing point.
}\label{s_max_fix}
\end{figure}

 Here we have considered the parameters to be energy independent. In this region, the maximum cross section are
\begin{eqnarray}
    \sigma^{\rm max}_{\omega\omega\to hh} &=& \frac{8\pi^3}{s} \, d_{\rm max}^2  \, \left(\frac{s}{16\pi^2 v^2}\right)^2 \,, 
\label{eq:max-cs2}
    \\
    \sigma^{\rm max}_{\omega\omega\to 3h} &=&  \frac{64\pi^3}{ 3s}   \, d_{\rm max}^4\, (1+\rho_{\rm max})^2 \,  \left(\frac{s}{16\pi^2 v^2}\right)^3  \, ,  
\label{eq:max-cs3}
    \\
    \sigma^{\rm max}_{\omega\omega\to 4h} &=&  \frac{8\pi^3}{9 s}\left(\frac{s}{16\pi^2 v^2}\right)^4  \, d_{\rm max}^4\,  \bigg[  (1+\rho_{\rm max})^2 + 2(1+\rho_{\rm max}) \chi_1    + \chi_2  
\bigg] \, ,
\label{eq:max-cs4} 
\end{eqnarray}
where the maximum is reached at any $\rho\in[-\rho_{\rm max}\, ,\, \rho_{\rm max}]$ for $2h$ and at $\rho=\rho_{\rm max}$         
for $3h$ and $4h$ (for $1+\chi_1\geq 0$).    
Figure~\ref{s_max_fix} shows these maximized cross sections as a function of the center-of-mass energy $\sqrt{s}$. At low energies there is a hierarchy in the multi-Higgs production cross sections, where $2h$ production is larger than $3h$, and this larger than $4h$. 
However, as the energy increases there is a moment in which all three cross sections become roughly of the same order and then the hierarchy gets inverted being, according to the theoretical expression, more likely to generate $4h$. This feature only highlights that from that point on, the EFT should not be trusted anymore, as we have reached the cut-off of the theory. At that point corrections from higher orders are as big as the contributions already included in~(\ref{cs2}), (\ref{cs3}) and (\ref{cs4}). 
Beyond that point, one should not further rely on the information provided by this plot. 
Nevertheless, the $\sqrt{s}\lsim 10^1$~TeV range in figure~\ref{s_max_fix} shows the exclusion plot for the multi-Higgs production cross section one can expect for the allowed SMEFT range in~(\ref{eq:allowed-d-rho}).

\begin{figure}[t!]   
    \centering
 {
  \includegraphics[width=0.8\textwidth]{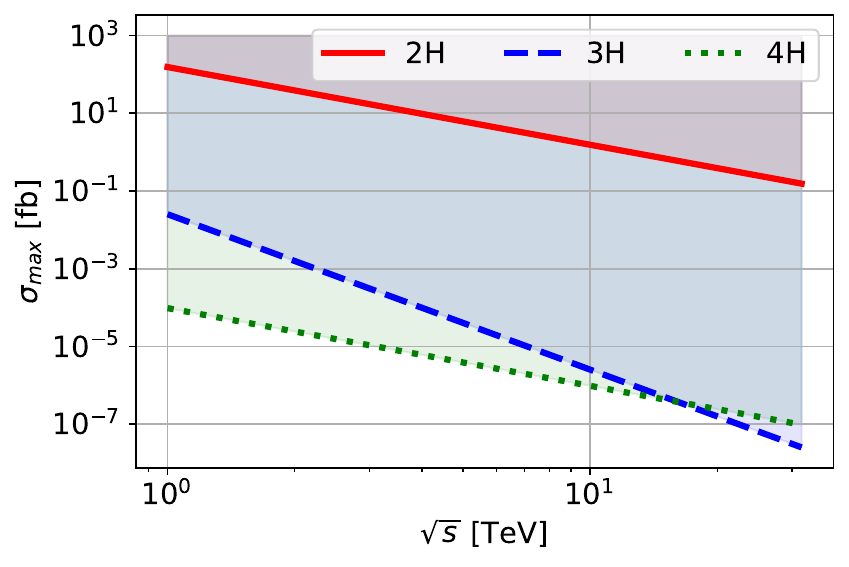}
  \vspace*{0.5cm}
  }
  \caption{{\small Exclusion plot for the maximum value of the cross sections for 2, 3 and 4 Higgs bosons with  
  the constraint $|\rho|\leq \rho_{\rm max}=1$ and EFT-expansion tolerance $\epsilon=0.1$.   }}
  \label{s_max_var}
\end{figure}

Figure~\ref{s_max_fix} may be a little bit unsatisfactory, as the high energy part of the plot lies off the EFT applicability region. Thus, one might want to study the EFT prediction for all possible energies whenever the EFT is still applicable. For this, at a given energy, one needs to establish a practical criterion on what ``EFT-valid'' means. For theories where one has $D=6$ contributions it seems fair to assume that one has an EFT expansion in powers of $\frac{c_{H\Box}^{(6)} s}{\Lambda^2} = \frac{d\, s}{2v^2}\ll 1$.  
Hence, for that given $\sqrt{s}$, we will allow all values for $d$ that obey, 
\begin{equation}
\left|\frac{c_{H\Box}^{(6)} s}{\Lambda^2}\right| \, = \, \left|\frac{d\, s}{2v^2}\right| \, \leq  \, \epsilon \, \ll 1 \, ,
\end{equation}
for some EFT-expansion tolerance parameter $\epsilon$ that controls the rate of convergence. This implies that for each value of the energy we are going to consider a range for $d$ that varies with $s$ in the form, 
\begin{equation}
|d|\, \leq d_{\rm max}(s) \, =\, \frac{2 v^2}{s}\, \epsilon\, . 
\end{equation}
This choice of the allowed range for $d$ ensures that the EFT analysis is valid at all considered energies. If now substitute this maximum value $d_{\rm max}=d_{\rm max}(s)$ in the previous cross section bounds~(\ref{eq:max-cs2})-(\ref{eq:max-cs4}), we obtain the modified expressions, 
\begin{eqnarray}
    \sigma^{\rm EFT-max}_{\omega\omega\to hh} &=& 
    \frac{\epsilon^2}{8\pi s}\,,
\label{eq:EFT-max-cs2}
    \\
    \sigma^{\rm EFT-max}_{\omega\omega\to 3h} &=&  
    \left(\frac{v^2}{16\pi^2 \, s}\right)\, \frac{4\epsilon^4}{3\pi s} \, (1+\rho_{\rm max})^2 \, ,
\label{eq:EFT-max-cs3}
    \\
    \sigma^{\rm EFT-max}_{\omega\omega\to 4h} &=&  
    \left(\frac{1}{16\pi^2}\right)^2\, \frac{\epsilon^4}{18\pi s} \, ( (1+\rho_{\rm max})^2 + 2(1+\rho_{\rm max}) \chi_1      +\chi_2)\, ,
\label{eq:EFT-max-cs4} 
\end{eqnarray}
which ensure the EFT validity at all explored energies. 
Figure~\ref{s_max_var} shows the exclusion regions for $2$, $3h$ and $4h$ cross sections for the bound $\rho_{\rm max}=1$ and the tolerance $\epsilon=0.1$.

\subsection{A collider estimate illustration: $e^+e^-$ CLIC at 3~TeV}

 \begin{figure}[t!] 
    \centering
 {
  \includegraphics[width=0.6\textwidth]{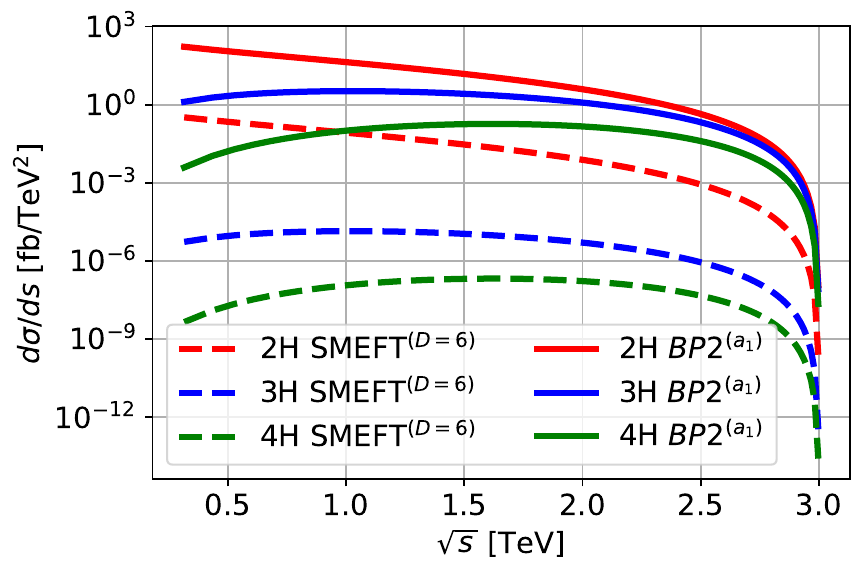}
  \vspace*{0.5cm}
  }
  \caption{{\small 
     $e^+e^-\to e^+e^- \, +\, n\times h$ differential cross section in EWA~ \cite{Dawson:1984gx,Cahn:1983ip,Chanowitz:1984ne,Kane:1984bb,Dobado:2015hha}.  We provide the $\rm{SMEFT}^{(D=6)}$ (dashed line) and $\rm{BP}2^{(a_1)}$ (solid line) benchmarks for illustration. 
  We show the cross sections for $2h$ (red), $3h$ (blue) and $4h$ (green) production. In general, for each BP one observes the ordering $\frac{d\sigma_{2h}}{ds}< \frac{d\sigma_{3h}}{ds} < \frac{d\sigma_{4h}}{ds}$.  } 
}\label{fig:diff_cs_ewa}
\end{figure}

In this last subsection we want to provide an illustration of the expected cross sections in an actual collider, not just for the hard subprocess $\omega\omega\to n\times h$. 
We do not intend, however, to perform a full detailed simulation of the expected result in a fully realistic collider. Its complexity relegates that analysis for a future work. In this article we are going to focus on the rough estimate provided by the {\it effective W~approximation} (EWA)~\cite{Dawson:1984gx,Cahn:1983ip,Chanowitz:1984ne,Kane:1984bb,Dobado:2015hha}, where the longitudinal gauge bosons are collinearly radiated from the incoming particles. This kinematical region is expected to dominate the full cross section and provide a reasonable estimate. 

Regarding the type of collider, for the sake of simplicity, we will consider a lepton collider such as CLIC ($e^+e^-$) with $\sqrt{s^{\rm tot}}=3$~TeV~\cite{Linssen:2012hp}.  
In Stage~3 scenario the integrated luminosity for that center-of-mass total energy may reach up to 5000~fb$^{-1}$~\cite{CLIC:2018fvx}. This roughly means that cross sections with $\sigma\lsim 0.2\cdot 10^{-3}$~fb are not expected to be observed. In practice, the lowest observable bound is actually even higher, as one needs to take into account the efficiency of the detectors in the reconstruction of the Higgs decays.

\begin{figure}[t!] 
    \centering
 {
  \includegraphics[width=0.8\textwidth]{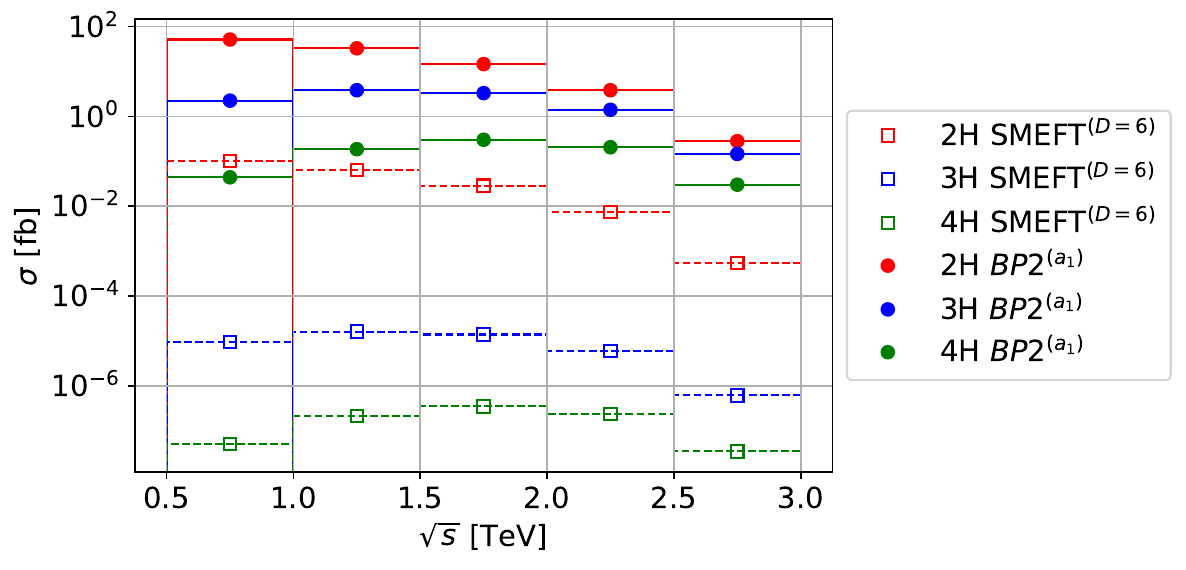}
  \vspace*{0.5cm}
  }
  \caption{{\small  $e^+e^-\to e^+e^- \, +\, n\times h$ cross section in the effective W approximation~ \cite{Dawson:1984gx,Cahn:1983ip,Chanowitz:1984ne,Kane:1984bb,Dobado:2015hha}. Note that in this plot the height of each bin provides the integrated cross section for the corresponding intervals $[\sqrt{s_k},\sqrt{s_{k+1}}]$, with $\sqrt{s_k}=k\times 0.5$~TeV and $k=1,2,3$, etc.   
  We provide the $\rm{SMEFT}^{(D=6)}$ (empty square and dashed line) and $\rm{BP}2^{(a_1)}$ (filled circle and solid line) benchmarks for illustration. 
  We show the cross sections for $2h$ (red), $3h$ (blue) and $4h$ (green) production. In general, for each BP one observes the ordering $\sigma_{2h}< \sigma_{3h} < \sigma_{4h}$. 
  }  
}\label{fig:cs_ewa}
\end{figure}

The EWA relates the total cross section for $e^+e^-\to e^+e^- \,+\, n\times h$ produced through an intermediate $W_L^+W_L^-$ pair and the $W_L^+W_L^-\to n\times h$ cross section. In the EWA factorization, the total cross section, $\sigma_{tot}$, is provided by the hard subprocess cross section times an appropriate $W_L$--luminosity function of the form~\cite{Dobado:2015hha}, 
\begin{equation}
\frac{d\sigma_{tot}}{ds} =\frac{\alpha^2}{8\pi^2  s\, \, \sin^2\theta_W}\, \, \bigg[2\left(\frac{s}{s^{\rm tot}}-1 \right)- \left(\frac{s}{s^{\rm tot}}+1\right) \log{\frac{s}{s^{\rm tot}}} \bigg]\,\times \, \sigma(s)\bigg|_{W_L^+W_L^-\to n\times h}\, ,    
\end{equation}
being $s=p_{W^+W^-}^2=p_{n\times h}^2$, $\alpha$ the fine-structure constant and $\theta_W$ the Weinberg angle.

This expression provides the dependence of the differential total cross section on the $WW$ invariant mass, $\sqrt{s}$, which has been plotted in Fig.~\ref{fig:diff_cs_ewa} for the production of 2, 3 and 4 final Higgs bosons. We are providing the predictions for the $\rm{SMEFT}^{(D=6)}$ benchmark (dashed lines) discussed in this and previous sections, considering $d=0.1$ and $\rho=1$. It is possible to observe the cross-section suppression for SMEFT scenarios as we increase the number of final Higgs bosons. For illustration we also provide the differential cross section for a pure-HEFT benchmark, $\rm{BP}2^{(a_1)}$. The comparison of the two scenarios shows again that, in general, pure-HEFT models lead to cross sections which are orders of magnitude larger than those in theories that admit a low-energy SMEFT description. We note that all EWA differential cross-sections vanish at the end-point $s=s^{\rm tot}$.

To present these results in a more transparent form, in Fig.~\ref{fig:cs_ewa}, we provide the integrated cross section over bins of $0.5$~TeV. For the pure-HEFT scenario $\rm{BP}2^{(a_1)}$ one finds cross sections ranging in the order from $10^{-2}$ to $10^2$~fb, which could be potentially observed with a 5000~fb$^{-1}$ integrated luminosity.
In the $\rm{SMEFT}^{(D=6)}$ benchmark this only happens for the two-Higgs  production: the production of 3 or more Higgs bosons shows cross sections well below the attobarn by several orders of magnitude.
In this sense, the observation of multi-Higgs states in that referred collider would indicate that the underlying UV theory could not support a SMEFT-like low-energy description. It should then be described through the more general HEFT formalism.

For illustration, we have selected a specific lepton collider, providing a clean and clear result. An analogous analysis can be performed for other lepton colliders, as well as for current or future hadron accelerators. However, extending this analysis to hadron colliders requires further discussion on quark parton distribution functions and, to make these predictions useful, a good estimation of detector, reconstruction, and analysis effects. Although very interesting for the proposals of future hadron colliders, such a study is beyond the scope of this work.

\section{Conclusions}
\label{sec:conclusions}

In this article we have performed an EFT analysis for longitudinal VBS into several Higgs bosons. In order to clarify the relevant interactions at energies well over the production threshold, we have employed the EqTh and considered a series of well-motivated approximations, such as assuming the mass corrections to be subdominant with respect to momentum dependent interactions. These approximations have allowed us to relate the VBS with the EW Goldstone scattering process $\omega\omega\to n\times h$.

In section~\ref{sec:HEFT-scat}, we have presented the calculations for $2h$, $3h$ and $4h$ production from $\omega\omega$ in the HEFT approach at its lowest order. This means we have considered only tree-level processes and contributions from HEFT operators with up to two derivatives, $\mO(p^2)$. Loops or operators with a higher number of derivatives lead to further subleading corrections, neglected in this article. 

We provide analytical expressions for the scattering amplitudes and cross sections, the only exception being the $\omega\omega\to 4h$ cross section which is provided through an analytical expression in terms of two coefficients $\chi_{1,2}$, which must be computed numerically. A code ({\tt MaMuPaXS}~\cite{MaMuPaXS}) has been specifically developed for the phase-space integration with the massless particle approximation assumed here.
This code greatly reduces the computational resources required to sample the cross section at different energies.

We want to note that we have identified the relevant combinations $\hat{a}_{2,3,4}$ that rule the $\omega\omega\to 2h,\, 3h,\, 4h$ scattering amplitudes. These are the relevant parameters one should aim for in future experimental analyses, and not the $WWh$, $WWh^2$ and $WWh^3$ couplings $a_{2,3,4}$, respectively. Note that, trying to measure such couplings experimentally leads to unwanted strong correlations in the data analyses.    
We have shown that through scalar field redefinitions in the effective action (changes of coordinates in the scalar manifold) one can derive these precise relevant combinations $\hat{a}_j$ (see appendix~\ref{app:field-redef}). Indeed, it is easy to extract, e.g., the relevant combinations $\hat{a}_5$ and $\hat{a}_6$ for $5h$ and $6h$ production even before starting the computation of this amplitude.

In section~\ref{sec:SMEFT-scat} we restricted those results to the case when the EFT accepts a SMEFT representation, this is, when one can write down the effective Lagrangian in terms of a Higgs doublet and sort out the operators according to their canonical dimension. It is obviously implicit that, in addition to the latter consideration, this SMEFT power counting in canonical dimension gives a convergent EFT expansion for the observables, which does not always need to be true for an arbitrary EFT. 
As before, we focus on a regime well above threshold where momentum dependent interactions dominate over non-derivative ones.  
We compute the contributions to the tree level $\omega\omega\to 2h,\, 3h,\, 4h$ amplitudes from operators of canonical dimensions $D=6$ and $D=8$. 
We provide the results for the amplitudes and total cross sections in compact analytical expressions. 
Remarkably, we find that under our approximations the multi-Higgs productions is highly suppressed in the SMEFT, since producing a large number of Higgs bosons requires the insertion of an increasing number of $D=6$ operators of terms of higher dimension ($D=8,\, 10,\,$etc.).

In section~\ref{sec:pheno} we have considered a series of benchmark points to compare scenarios where the effective theory accepts a SMEFT description and pure-HEFT scenarios, where the theory cannot be written in the SMEFT form. 
The study of VBS into $2h$, $3h$ and $4h$ shows that in general the non-SMEFT cross sections are orders of magnitude larger than the SMEFT predictions. This happens even if we set the $WWh$ and $WWh^2$ couplings identical to the SMEFT ones. Moreover, in the non-SMEFT scenarios the cross-section values do not dramatically depend on the precise values chosen for the $a_j$ and they are stable under small modifications of these parameters. By contrast, SMEFT highly rely on very fine-tuned correlations between the $WWh^n$ couplings, which lead to very precise cancellations in the cross sections that drop their value by several orders of magnitude. 

Although in some cases, the amplitude for a specific number of Higgs bosons can be abnormally small in a particular model (e.g., $\omega\omega\to 2h$ for BP1$^{(a_1)}$), in general for any other number of final Higgs bosons non-SMEFT cross sections are orders of magnitude larger than the SMEFT predictions. 
Thus, a comparative analysis of VBS into channels with different numbers of Higgs bosons can tell us whether the (non-resonant) experimental data can be described through SMEFT. Indeed, section~\ref{sec:SMEFT-exclusion} shows the exclusion cross section plots for $\omega\omega\to 2h,\, 3h,\, 4h$ if one assumes SMEFT: as it does not allow large cross sections in multi-Higgs production one finds that the maximum allowed SMEFT cross section is much more suppressed than the pure-HEFT one. For our BP's we found SMEFT yields cross sections below the attobarn for $\sqrt{s}\gsim 2$~TeV for VBS into more than two Higgs bosons (see figure~\ref{s_max_var}).  For illustration, we have studied the multi-Higgs production cross section in a 3~TeV $e^+e^-$ collider (CLIC) within the EWA, where the longitudinal $W^+_LW^-_L$ are radiated collinearly from the incoming $e^+e^-$. The cross section estimates in Fig.~\ref{fig:cs_ewa} lead to the same conclusions found in previous sections: multi-Higgs production is strongly suppressed in SMEFT scenarios in comparison with pure-HEFT cases.

In summary, it is complicated to produce large deviations from the SM in multi-Higgs VBS within the SMEFT framework. An eventual experimental observation of cross sections that exceed the bounds prescribed by SMEFT (Figs.~\ref{s_max_fix} and~\ref{s_max_var}) would be an indication that a broader approach is needed. Thus, the joint study of the $2h$, $3h$ and $4h$ production channels can serve as a smoking gun indicating that (non-resonant) new physics, if present, requires an interpretation beyond SMEFT, requesting the use of the full HEFT machinery. We are confident that this work can serve as a motivation for studies in future colliders, illustrating what possible (non-resonant) new physics could we find.   

There are several directions that can be followed as next steps. 
First, one should go beyond the EqTh approximation and consider mass corrections. This would allow us to also study the region close to threshold and improve the accuracy of our high energy EFT predictions. 
In addition, the cross sections studied here referred to the hard subprocesses $WW\to n\times h$ while in colliders, either leptonic or hadronic, the $W$'s need to be radiated from the colliding particles, diminishing the total $n\times h$ production cross section for the VBS category. Likewise, simplifications as the effective $W$ approximation may not 
be fully satisfactory and a full MonteCarlo simulation may be required for a precise LHC estimate. One may find then that the signal studied in this article may appear surrounded by additional reducible and irreducible background that needs to be disentangled. But nevertheless, the signal is there and it should give some definite prediction for $pp$ or $e^+e^-$ colliders if the experimental data and MonteCarlo simulations are improved enough. 
A final obvious improvement is the consideration of next-to-leading order and loop corrections in the EFT~\cite{Herrero:2022krh,Herrero:2021iqt,Quezada-Calonge:2022lop}, 
which would further refine the predictions.

\acknowledgments

We thank C. Quezada for useful checks on the {\tt MaMuPaXS} phase-space integration. We are grateful to S. Badger, A. Dobado and F. J. Llanes-Estrada for their careful reading and comments on the manuscript. We also thank F. Arco, M. Cepeda, M.J. Herrero, J. Le\'on, N. Smith and J.F. de Troc\'oniz for useful discussions on multi-Higgs production at the LHC.  \    
This work has been supported in part by Spanish MICINN (PID2022-137003NB-I00,\ \    PID2021-124473NB-I00,   \ \ PID2019-108655GB-I00/AEI/10.13039/501100011033), U. Complutense de Madrid under research group 910309, the IPARCOS institute, the EU under grant 824093 (STRONG2020). 
The work of RGA is supported by the EU’s Next Generation grant DataSMEFT23 (PNRR - DM 247 08/22). The work of JMM is supported by the grant Ayudas de doctorado IPARCOS-UCM/2022. ASB acknowledges the support of the EU's Next Generation funding, grant number CNS2022-135688.
RGA thanks the Galileo Galilei Institute for hospitality and support during the scientific program on “Theory Challenges in the Precision Era of the Large Hadron Collider,” where part of this work was carried out. 

\appendix

\section{Phase-space relations}
\label{app:phase-space}

The Lorentz invariant phase-space for $n$ final particles is given by 
\begin{equation}
d\Pi_n \,=\, \displaystyle{ 
(2 \pi)^4 \, \delta^{(4)}\left(\sum_{i=1}^n p_i\,-\, q\right)\, \prod_{j=1}^n \frac{d^3\vec{p}_j}{(2\pi)^3\, 2E_j} \, ,
}
\label{eq:phasespace}
\end{equation}
with $E_j={\vec{p}_j}^{\,2}+m_j^2$.  

In the massless case ($m_j=0$) one has $E_j=|\vec{p}_j|$ and $d^3\vec{p}_j/E_j= |\vec{p}_j|\,  d|\vec{p}_j|\, d\Omega_j^\ast$.  
It is then possible to provide an analytical expression for the full phase-space hypervolume~\cite{Salas-Bernardez:2020hua}:
\begin{equation}
\mathcal{V}_n\, =\, \displaystyle{\int d\Pi_n}\, =\, \frac{1}{2(4\pi)^{2n-3}}\frac{s^{n-2}}{\Gamma(n)\Gamma(n-1)}\, .  
\label{eq:phasespaceintegr}
\end{equation}

It will be convenient in this massless case, for the formal understanding of the energy dependence and the numerical simulations, to factorize out the global $s^m$ dependence of the phase-space integral through the change of variables $\widetilde{p}_i =p_i/\sqrt{s}$ 
and $\widetilde{q}=q/\sqrt{s}$, defining:
\begin{eqnarray}
\label{eq:phasespace-tilde} 
d\widetilde{\Pi}_n &=& s^{2-n}\, d\Pi_n  
\,=\, \displaystyle{ 
(2 \pi)^4 \, \delta^{(4)}\left(\sum_{i=1}^n \widetilde{p}_i\, -\, \widetilde{q}\right)\, \prod_{j=1}^n \frac{d^3\vec{\widetilde{p}}_j}{(2\pi)^3\, 2\widetilde{E}_j}  }
\, , 
\\
\label{eq:phasespaceintegr-tilde}
\widetilde{\mathcal{V}}_n &=&  \displaystyle{\int d\widetilde{\Pi}_n}\, =\, \frac{1}{2(4\pi)^{2n-3} \Gamma(n)\Gamma(n-1)}\, , 
\end{eqnarray}
with the delta imposing the energy conservation relation $\sum_i \widetilde{E}_i=1$ and, for the center-of-mass frame, the three-momentum conservation $\sum_i \widetilde{\vec{p}}_i=0$. 
For the simplest case, $n=2$, this expression recovers the standard result $\mathcal{V}_2=\widetilde{\mathcal{V}}_2=\int d\Pi_2=\mathcal{V}_2=(8\pi)^{-1}$. 
The advantage of this $s$-independent phase-space is that the integration of an arbitrary phase-space of $n$ particles will always handle $\mO(1)$ numbers and trigonometric functions. Thus, if one can factorize out the $s$ dependence of the scattering amplitude, one does not need to repeat the $\int d\Pi_n\,\left(\diamond\right)$ phase-space integration for every energy. 

\section{4-momentum conservation: kinematical relations} \label{app:mom-conservation}

Given a process $\omega(k_1)\, \omega(k_2)\to h(p_1)... h(p_n)$. In the rest frame, the center of mass squared energy, $s=(k_1+k_2)^2$, equals
\begin{equation}
    s\equiv 4\lVert\vec{k}_1\rVert^2 \,. 
\end{equation}
We define the rest-frame three-momentum fractions as
\begin{equation}
f_i\equiv\lVert\vec{p}_i\rVert/\sqrt{s} \,, 
\end{equation}
for each outgoing Higgs boson; the functions
\begin{equation}
z_i\equiv 2\sin^2(\theta_i/2) = 1-\cos(\theta_i) \,, 
\end{equation}
where $\theta_i$ is the angle between the $i$-th Higgs boson and the first $\omega$ Goldstone boson momentum, ${\vec{k}}_1$ (that is, $z_1=1-\cos\theta$, $z_2=1+\cos\theta$ as usual in a two-body problem with $t$ and $u$ channels); and the functions
\begin{equation}
z_{ij}\equiv 2\sin^2(\theta_{ij}/2)= 1-\cos(\theta_{ij})\, , 
\end{equation}
where $\theta_{ij}$ is the angle between the $i$-th and $j$-th Higgs bosons.

Taking the masses of both the Golstones and Higgses as zero, we have the 4-momentum products:
\begin{eqnarray}
&&   k_1p_i=\frac{s}{2}f_iz_i \, ,
\\
&&  p_ip_j=sf_if_jz_{ij}\, ,
\end{eqnarray} 
where no implicit summation is assumed in repeated indices. These expressions allow us to write down these functions in terms of Lorentz invariant products:
\begin{equation}
f_i = \frac{q p_i}{q^2} \, , \qquad 
z_i = \frac{2 k_1 p_i}{q p_i}\,, \qquad 
z_{ij} = z_{ji}= \frac{q^2\, (p_ip_j)}{(q p_i)\, (q p_j)}\, ,
\end{equation}
with $q\equiv k_1+k_2$. If one wants the amplitudes in the main text in terms of Lorentz invariant scalar products one just needs to perform the replacements above.

Using 4-momentum conservation, we can obtain some relations between the newly defined functions. Firstly, we have  $\left(\displaystyle\sum_{i=1}^n {p_i}\right)\cdot\left(\displaystyle\sum_{i=1}^n p_i\right)=s$,     
which is equivalent to
\begin{equation}
\sum_{\substack{i,j=1 \\ i<j}}^nf_if_jz_{ij}=\frac{1}{2} \,.
\end{equation}

Also, from conservation of the total energy in the rest frame, we get $\displaystyle\sum_{i=1}^np_i^0=\sqrt{s}$\, which is equivalent to
\begin{equation} \label{rel:fff1}
\sum_{i=1}^nf_i=1 \,.
\end{equation}

Then, using ${k_1}\cdot\left(\displaystyle\sum_{i=1}^n{p_i}\right)=s/2$, we get
\begin{equation}
\sum_{i=1}^nf_iz_i=1 \,.
\end{equation}

Furthermore, using \,$\vec{p}_j\cdot\left(\displaystyle\sum_{i=1}^n\vec{p}_i\right)=0$ and \eqref{rel:fff1}, we have
\begin{equation}
\sum_{\substack{i=1 \\ i\neq j}}^n f_iz_{ij} = 1 \quad \forall j\in\{1,...,n\} \text{ with } f_j\neq 0 \, . 
\end{equation}

Lastly, using that $\lVert\vec{p_j}\rVert^2=\left(\displaystyle\sum_{\substack{i=1 \\ i\neq j}}^n\vec{p_i}\right)^2$ \, for all $j\in\{1,...,n\}$ and \eqref{rel:fff1}, we obtain

\begin{equation}
\frac{1}{2}-\sum_{\substack{i,j=1 \\ i<j \\ i\neq k \\ j\neq k}}^nf_if_jz_{ij} = f_k \quad \forall k\in\{1,...,n\} \, . 
\end{equation}

As an example, for three Higgs bosons in the final state, the relevant relations are
\begin{eqnarray}
&&   f_1 f_2 z_{12} +  f_1 f_3 z_{13} +  f_2 f_3z_{23}= \frac{1}{2}  \, ,   
\\
&&  f_1z_1+f_2z_2+f_3 z_3 \, =\, 1 \, \text{ with } f_i\neq 0\,, 
 \\
&& f_1+f_2+f_3=1\, . 
\end{eqnarray}   

Furthermore, we can use this relations to write $B$ of eq.~(\ref{eq:B}) only in terms of $f_{1,2,3}$, $z_{1,2,3}$,  $z_{12}$ and $z_{13}$ as
\begin{eqnarray} 
\label{eq:B2}
B &=& \frac{f_1f_2z_{12}\left(f_1f_2z_{12}-f_1-f_2+\frac{1}{2}\right)}{2f_1f_2z_{12}-f_1z_1-f_2z_2} + \frac{f_1f_2z_{12}\left(f_1f_2z_{12}-f_1-f_2+\frac{1}{2}\right)}{2f_1f_2z_{12}-2f_1-2f_2+f_1z_1+f_2z_2} + \nonumber\\ 
&&+ \frac{f_1f_3z_{13}\left(f_1f_3z_{13}-f_1-f_3+\frac{1}{2}\right)}{2f_1f_3z_{13}-f_1z_1-f_3z_3} + \frac{f_1f_3z_{13}\left(f_1f_3z_{13}-f_1-f_3+\frac{1}{2}\right)}{2f_1f_3z_{13}-2f_1-2f_3+f_1z_1+f_3z_3} + \nonumber\\ 
&&+ \frac{\left(f_1f_2z_{12}+f_1f_3z_{13}-f_1\right)\left(f_1f_2z_{12}+f_1f_3z_{13}-f_1-f_2-f_3+\frac{1}{2}\right)}{-2f_1f_2z_{12}-2f_1f_3z_{13}+2f_1+f_2z_2+f_3z_3-1} + \nonumber\\
&&+ \frac{\left(f_1f_2z_{12}+f_1f_3z_{13}-f_1\right)\left(f_1f_2z_{12}+f_1f_3z_{13}-f_1-f_2-f_3+\frac{1}{2}\right)}{-2f_1f_2z_{12}-2f_1f_3z_{13}+2f_1+2f_2+2f_3-f_2z_2-f_3z_3-1} \, .
\end{eqnarray}    
This expression for $B$ has been checked analytically with Maple and numerically with the integration code in ref.~\cite{MaMuPaXS}.

\section{HEFT simplifications through field redefinitions:\texorpdfstring{\\}{}
further checks and understanding of \texorpdfstring{$\omega\omega\to n \times h$}{ww->nxh}:}
\label{app:field-redef}

As stated in the main text, we have used the standard form of the HEFT Lagrangian~(\ref{eq:HEFT-Lagr}) to compute every possible tree-level Feynman diagram contributing to the processes discussed in this article. We have performed the calculation within the equivalence theorem and neglected non-derivative interactions, keeping only the $\mO(E^2)$ contributions to the amplitudes. 

After some algebra, we have been able to simplify the amplitudes, finding that both for $2h$ and $3h$ production the amplitudes are pure $s$-waves. The crossed-channel Goldstone exchanges cancel out and we are effectively left with a contact interaction (in spite of the large variety of crossed channel diagrams). For the case with $4h$ in the final state, we also find strong cancellations and the amplitude is effectively reduced to a contact $\omega\omega\to 4h$ interaction plus a contribution with one Goldstone exchange in the crossed channel.

In this appendix we proceed to clarify the origin of such cancellations, which lead to the extremely simple structure of our results. 
The part of the HEFT Lagrangian~(\ref{eq:HEFT-Lagr}) relevant for the $\omega\omega\to n\times h$ scattering has the form,
\begin{equation}    
\mL =\frac{1}{2}\partial_\mu h \partial^\mu h 
    + \frac{1}{2}  \mF(h)\, \partial_\mu\omega^a\partial^\mu\omega^a   \, \,+\, \mO(\omega^4) \, ,  
\label{eq:simpler-L-scat}
\end{equation}
with the flare function $\mF(h)=1+ a_1 h/v + a_2 h^2/v^2 +a_3 h^3/v^2+a_4 h^4/v^2 + \mO(h^5)$.

As an alternative to this canonical HEFT Lagrangian~(\ref{eq:HEFT-Lagr}), 
we realized that these processes can be simplified by performing a non-linear field redefinition~\cite{Criado:2018sdb,Chisholm:1961tha,Kamefuchi:1961sb,Divakaran:1963yxz,Arzt:1993gz} in~(\ref{eq:HEFT-Lagr}) before starting the amplitude computation. The aim of these redefinitions is to simplify and eliminate the derivative three-particle vertex $h\omega\omega$~\cite{Georgi:1991ch,Xiao:2007pu}. In this way, the number of possible diagrams for a given process is greatly reduced~\footnote{This absence of 3-particle vertices might be convenient even for loop-calculations: this type of interactions induces tadpole diagrams at one-loop, which require a readjustment of the vacuum at that order. Nonetheless, this field redefinition may affect different sectors, such as the Yukawa one, complicating the amplitudes therein.}. 

For this simplification we have considered transformations of the form, 
\begin{eqnarray}\label{eq:C2}
&&   \omega^a\to \omega^a 
+ g(h) \, \omega^a 
\,,
\qquad    
h \to h 
+ \mathcal{N} \, (1+ g(h)) \, \omega^a\omega^a /v
\,, 
\end{eqnarray} 
with the free dimensionless real constant $\mN$ and the $\mO(h)$ function $g(h)$ determined by the flare function $\mF(h)$ through the relation $ g'(h)=- 2\mN/[ v\, \mF(h)]$, leading to 
\begin{equation}
\displaystyle{    g(h)\, =\, -\, \frac{2 \mN}{v}\,  \int_0^h \frac{ds}{ \mF(s)  }     } 
\,= \mN\, \bigg( \, -2 \frac{h}{v} + 2 a \frac{h^2}{v^2} + \frac{2}{3} (b-4a^2) \frac{h^3}{v^3} + \frac{1}{2} (a_3 -4 ab  + 8 a^3) \frac{h^4}{v^4} +\mO(h^5)\bigg)  \, .    
\label{eq:g-transform}
\end{equation} 
Interestingly, the application of this transformation to the Lagrangian~(\ref{eq:simpler-L-scat}) leads to a new Lagrangian with exactly the same structure in~(\ref{eq:simpler-L-scat}), but with a new function $\hat{\mF}(h)$ determined by: 
\begin{equation}
\hat{\mF}(h)\, =\, \mF(h)\, \big(1+g(h)\big)^2. 
\label{eq:Fhat}
\end{equation}

In particular, we will consider here a very specific $\mN$ normalization: 
\begin{equation}
\mN\,=\, \frac{a}{2}\,, 
\qquad 
g(h)\,=\, -a \frac{h}{v} +a^2 \frac{h^2}{v^2} +\frac{1}{3}a(b-4 a^2) \frac{h^3}{v^3}  + \frac{1}{4} a( a_3 - 4 ab + 8 a^3) \frac{h^4}{v^4} \, +\, \mO(h^5)\, .
\label{eq:g-transform-a1=0}
\end{equation}
This choice of the scalar manifold coordinates transforms the original Lagrangian~(\ref{eq:simpler-L-scat}) into 
\begin{eqnarray}    
\mL &=& \frac{1}{2}\partial_\mu h \partial^\mu h 
    + \frac{1}{2} \hat{\mF}(h)\, \partial_\mu\omega^a\partial^\mu\omega^a   \, \,+\, \mO(\omega^4)\, 
    \, ,  
\label{eq:simpler-L}
\end{eqnarray}       
but now in terms of the new function,  
\begin{equation}
\hat{\mF}(h)\,=\, 1 
   + \hat{a}_2 \frac{h^2}{v^2} 
   + \hat{a}_3 \frac{h^3}{v^3} 
   + \hat{a}_4 \frac{h^4}{v^4} +\,\mO(h^5) \, ,
\end{equation}
with coefficients provided by the ones of $\mF(h)$ through the combinations
\begin{eqnarray}
\hat{a}_2 &=& a_2-\frac{a_1^2}{4}\,=\, b-a^2 \, , \qquad  
\\
\hat{a}_3 &=& a_3- \frac{2}{3}a_1 \left(a_2 -  a_1^2/4\right) \,=\,  a_3-\frac{4a}{3}\left(b-a^2\right)\,, \qquad 
\\
\hat{a}_4&=& a_4 -   \frac{3}{4} a_1 a_3    +   \frac{5}{12}a_1^2\left(a_2-a_1^2/4\right) 
\,=\, a_4 -   \frac{3}{2} a\,  a_3    +   \frac{5}{3}a^2\left(b-a^2\right) \, .
\end{eqnarray}
Note that $\hat{\mF}(h)$ has now its first non-trivial contribution at $\mO(h^2)$, in comparison with $\mF(h)$, where its first non-trivial term starts at $\mO(h)$.
We remind that both the $\mO(h^5)$ and $\mO(\omega^4)$ operators are irrelevant for the tree-level $\omega\omega$ scattering producing two, three and four Higgs bosons in the final state. 
Nonetheless, it is very simple to extract the relevant parameters for any $\omega\omega\to n \times h$ scattering: one just needs to extract $g(h)$ from~(\ref{eq:g-transform}), taking $\mN=a/2$ and including the power series terms up to order $h^n$, the latter included. One then substitutes this $g(h)$ in~(\ref{eq:Fhat}) to obtain $\hat{\mF}(h)$ up to that order.

Although this work does not consider production of a higher number of Higgs bosons, the procedure above allows us to extract the relevant combination of the flare function coefficients, $a_j$,  for a generic amplitude $\omega\omega\to n\times h$: 
\textbf{1st)} compute $g(h)$ up to $\mO(h^n)$ by plugging $\mF(h)$ in~(\ref{eq:g-transform}) up to that order with $\mN=a_1/4$; 
and \textbf{2nd)}  expand $\mF(h)=\hat{\mF}(h)\,\left(1+g(h)\right)^2$ up to $h^n$. The corresponding coefficients $\hat{a}_j$ will be the relevant combinations for that process. Data studies that do not take this into account and use directly the $a_j$ instead of the $\hat{a}_j$ essentially introduce unnecessary correlations that make the experimental analysis much more involved. 
As an example, though irrelevant for this work, we show that one can easily extract the next $\hat{a}_j$ coefficients:  
$\hat{a}_5= a_5-\frac{1}{120} a_1 \left(15 a_1 \hat{a}_3-16     \hat{a}_2^2+96 \hat{a}_4  \right)$,  
$\hat{a}_6=a_6+\frac{1}{180} a_1 \left(  7 a_1 \hat{a}_2^2  -27a_1    \hat{a}_4+45\hat{a}_2 \hat{a}_3-150 \hat{a}_5 \right)$, etc.

It is also interesting to note that in the dilatonic model~\cite{Halyo:1991pc,Goldberger:2007zk,Vecchi:2010gj,Chacko:2012sy,Bellazzini:2012vz}, where $\mF(h)=\left(1+ a h/v\right)^2$, one finds for the transformation above that $1+ g(h)=\left(1+ a h/v\right)^{-2}$, leading to $\hat{\mF}(h)=1$. So that, remarkably, all the $\hat{a}_j$'s couplings are zero and all $\omega\omega\to n\times h$ amplitudes vanish at tree level (within the EqTh limit considered in this article). 
This exact same conclusions apply to the SM, where $a=1$.        

Given that the generating functional of the quantum field theory is invariant under field redefinitions, both (\ref{eq:HEFT-Lagr}) and~(\ref{eq:simpler-L}) Lagrangians produce the same on-shell scattering amplitudes~\cite{Criado:2018sdb,Chisholm:1961tha,Kamefuchi:1961sb,Divakaran:1963yxz,Arzt:1993gz}. 
Nonetheless, we have explicitly checked that the Lagrangian in (\ref{eq:simpler-L}) exactly  reproduces the results in the text for 2, 3 and 4 Higgs production. 

\begin{figure}[!t] 
     \centering
     \begin{subfigure}[]
         \centering
         \begin{tikzpicture}[scale=1]
     \draw[dashed] (-1,1) -- (0,0)-- (-1,-1);
     \draw[] (0,0)-- (1,1);
       \draw[] (0,0)-- (1,-1);
       \draw[] (-1.25,1) node {$\omega$};
       \draw[] (-1.25,-1) node {$\omega$};
     \draw[] (1.25,1) node {$h$};
          \draw[] (1.25,-1) node {$h$}; 
\end{tikzpicture}
     \end{subfigure}
     \qquad
     \begin{subfigure}[]
         \centering
         \begin{tikzpicture}[scale=1]
     \draw[dashed] (-1,1) -- (0,0)-- (-1,-1);
     \draw[] (0,0)-- (1,1);
     \draw[] (0,0)-- (1,0);
       \draw[] (0,0)-- (1,-1);
       \draw[] (-1.25,1) node {$\omega$};
       \draw[] (-1.25,-1) node {$\omega$};
     \draw[] (1.25,1) node {$h$};
          \draw[] (1.25,-1) node {$h$}; 
           \draw[] (1.25,0) node {$h$};
\end{tikzpicture}
     \end{subfigure}
    \qquad
     \begin{subfigure}[]
         \centering
         \begin{tikzpicture}[scale=1]
     \draw[dashed] (-1,1) -- (0,0)-- (-1,-1);
     \draw[] (0,0)-- (1,1);
     \draw[] (0,0)-- (1,0.35);
     \draw[] (0,-0)-- (1,-0.35);
       \draw[] (0,0)-- (1,-1);
       \draw[] (-1.25,1) node {$\omega$};
       \draw[] (-1.25,-1) node {$\omega$};
     \draw[] (1.25,1) node {$h$};
     \draw[] (1.25,-0.35) node {$h$};
     \draw[] (1.25,0.35) node {$h$};
          \draw[] (1.25,-1) node {$h$}; 
\end{tikzpicture}
     \end{subfigure}
     \qquad
     \begin{subfigure}[]
         \centering
         \begin{tikzpicture}[scale=1]
     \draw[dashed] (-1,1) -- (0,1)--(0,-1)-- (-1,-1);
     \draw[] (0,1)-- (1,1);
     \draw[] (0,1)-- (1,0.35);
     \draw[] (0,-1)-- (1,-0.35);
       \draw[] (0,-1)-- (1,-1);
       \draw[] (-1.25,1) node {$\omega$};
       \draw[] (-1.25,-1) node {$\omega$};
     \draw[] (1.25,1) node {$h$};
     \draw[] (1.25,-0.35) node {$h$};
     \draw[] (1.25,0.35) node {$h$};
          \draw[] (1.25,-1) node {$h$}; 
\end{tikzpicture}
     \end{subfigure}
        \caption{\small {\bf a)} Only diagram contributing to the 
        process $\omega\omega\to 2h$. {\bf b)} Only diagram contributing to the  
        process $\omega\omega\to 3h$. {\bf c-d)} Only two diagrams contributing to the  
        process $\omega\omega\to 4h$. 
        We have used the simplified Lagrangian~(\ref{eq:simpler-L}) to generate these amplitudes, so every $\omega\omega h^n$ vertex carries an $\hat{a}_n$ effective coupling. 
        Note that, in addition, one needs to consider all possible permutations for the assignment of the external particles. 
        }\label{diagrams}
\end{figure}
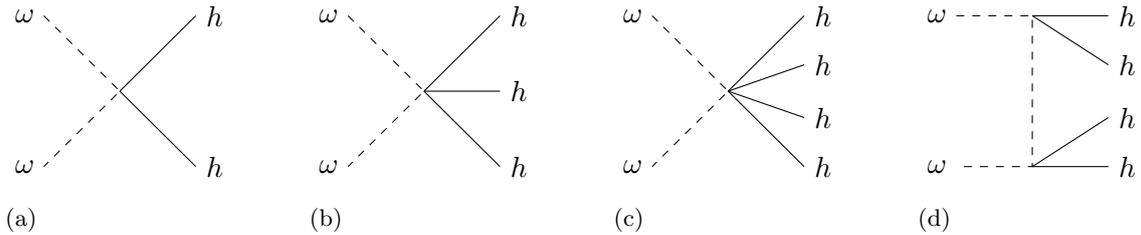

In this work we are neglecting the potential $V(h)$ (since $E\gg m_h$) and normalizing any pure-Higgs derivative operator of the form $f(h)\, (\partial h)^2$ into the canonical kinetic form $\frac{1}{2}(\partial h)^2$. Thus, there is no Higgs self-interaction and all the vertices relevant for the processes discussed in this article contain two derivatives and are of the type $\omega\omega h^n$. The critical gain from this field redefinition approach --using the simplified Lagrangian~(\ref{eq:simpler-L})-- 
is that the number of diagram topologies is greatly reduced:  
there is only 1 diagram for $\omega\omega\to 2h$ and $\omega\omega\to 3h$, and it is reduced to 2 diagram topologies for $\omega\omega\to 4h$ process (up to $n!$ permutations in the labeling of the outgoing Higgs particles in the diagrams). 
Figure~\ref{diagrams} shows all the relevant diagrams generated by~(\ref{eq:simpler-L}) for $\omega\omega\to 2h,\, 3h,\, 4h$, where in every $\omega\omega h^n$ vertex one now has the $\hat{a}_n$ effective coupling.

A closer observation of the transformation above shows that, in some scenarios, one can also conveniently choose the normalization $\mN$ to remove a higher order term, $a_n h^n$, from  \ $\mF(h)$ instead of the first.  For instance, provided ${ a_2< a_1^2/4 }$,  the choice \  $\mN= \left[ a_1 \pm\sqrt{a_1^2 - 4 a_2}\right]/4$ removes the $a_2 h^2$ term in $\mF(h)$, passing this information to the terms of order $h^1$ and $h^3$, $h^4$, etc. 

Another example is provided by the normalization $\mN=\frac{3}{8}\frac{a_3}{a_2-a_1^2/4}$, which removes the $h^3$ term in $\mF(h)$ and encodes its information in the factors $\hat{a}_j$ now multiplying $h^1$, $h^2$ and $h^4$, $h^5$, etc. This detail can be important for a proper interpretation of $WW\to 3h$ computations: at high energies, in the EqTh, it is possible to describe the $\omega\omega\to 3h$ scattering without an $\omega\omega h^3$ vertex ($\hat{a}_3=0$)~\cite{Gonzalez-Lopez:2020lpd},  
understanding that the $\omega\omega h^2$ and $\omega\omega h$ couplings are not the original ones ($a_2$ and $a_1$) but the effective ones in~(\ref{eq:simpler-L}) ($\hat{a}_2$ and $\hat{a}_1$). 

The main inconvenience of this approach is that the $SU(2)_L\times SU(2)_R$ chiral invariance of the action is no longer explicit as in~(\ref{eq:HEFT-Lagr}). The symmetry transformations are more involved in this form.  
For this reason, the correlations between the $a_1$ and $a_2$ couplings in $\mF(h)$ one finds for SMEFT-type theories with $D=6$ contributions 
(this is, $a_2=2a_1-3$~\cite{Gomez-Ambrosio:2022qsi,Gomez-Ambrosio:2022why})   
are no longer applicable for $\hat{\mF}(h)$ (SMEFT with $D=6$ contributions 
 does not fulfill $\hat{a}_2=2\hat{a}_1-3$, as one can see using, e.g.,~(\ref{eq:SMEFT-ajhat}) and $\hat{a}_1=0$);   
the chiral operator structures considered in~\cite{Gomez-Ambrosio:2022qsi,Gomez-Ambrosio:2022why} are deformed here for the terms of order $\omega^4$ and higher, so the conclusions therein are no longer applicable for the present simplified Lagrangian with $\hat{\mF}(h)$.      

Aside from these considerations, this type of simplifications may prove useful for the computation of EW processes with Higgs and Goldstone bosons in the equivalence theorem, either at tree or loop levels.

\section{Multi-Higgs production suppression in SMEFT: further clarifications}
\label{app:SMEFT-multiH-sup}

In order to obtain the valid tree-level configurations for a fixed number $n$ of final Higgses in $\omega\omega\to n \times h$, one has to vary $N_{V_3}$ from $0$ to $N_{V_3}^{\rm max}=n$, 
excluding the cases with $N_{V_4}=(n-N_{V_3})/2\notin\mathbb{N}\cup\{0\}$. 
It is illustrative to observe the valid diagrams for the cases with $n=2,3,4,5$ Higgses in the final state:
\begin{itemize}

    \item $n=2$ allows only diagrams with: $N_{V_3}=0, \, N_{V_4}=1$ ($N_I=0$);  $N_{V_3}=2, \, N_{V_4}=0$ ($N_I=1$).
    
    \item $n=3$ allows only diagrams with: $N_{V_3}=1, \, N_{V_4}=1$ ($N_I=1$);  $N_{V_3}=3, \, N_{V_4}=0$ ($N_I=2$). 
    
    \item $n=4$ allows only diagrams with: $N_{V_3}=0, \, N_{V_4}=2$ ($N_I=1$);  $N_{V_3}=2, \, N_{V_4}=1$ ($N_I=2$);  $N_{V_3}=4, \, N_{V_4}=0$ ($N_I=3$). 
    
    \item $n=5$ allows only diagrams with: $N_{V_3}=1, \, N_{V_4}=2$ ($N_I=2$);  $N_{V_3}=3, \, N_{V_4}=1$ ($N_I=3$);  $N_{V_3}=5, \, N_{V_4}=0$ ($N_I=4$). 
\end{itemize}

\section{Massless Multiparticle Cross Section ({\tt MaMuPaXS})}
\label{Appendix:MaMuPaXS}

\subsection{General case with 4 or more particles in the final state}
In order to numerically integrate the dimensionless function $B$ appearing in $\omega\omega\to 4h$ in eq.~(\ref{eq:B}), we are using the publicly available code \texttt{MaMuPaXS}~\cite{MaMuPaXS}. The code is based on Fortran, Python and the VEGAS library~\cite{Lepage:1977sw, Lepage:2020tgj} and can be accessed through a GitHub repository~\cite{MaMuPaXS}. 

For the integration of $B^n$ from (\ref{eq:chi-n_def}) our goal was keeping the computation as simple as possible. In order to achieve this, the change of variables described below (the {\tt MaMuPaXS} package) is a straightforward procedure for integrating the phase space if all the particles can be considered massless. Indeed, the only external library we use is VEGAS~\cite{Lepage:1977sw, Lepage:2020tgj}. However, there is the option of the RAMBO algorithm~\cite{Kleiss:1985gy}, a classic Monte-Carlo algorithm for generating phase space points and scattering events for multi-particle processes in the non-massless case. The question arising here is
whether, in the massless case, {\tt MaMuPaXS} (based on solving a $4\times 4$ system of linear equations and computing a $4\times 4$ determinant --the Jacobian) outperforms RAMBO. But, since we were not bounded by CPU time, this was a collateral issue. Hence, we have currently implemented up to $2\to 5$ processes on {\tt MaMuPaXS} 
(though this article only studies up to $2\to 4$ processes). 
However, an extension to arbitrary $2\to n$ processes is feasible and, indeed, should be made in order to compare with RAMBO's performance. This is a desirable goal for a continuation of the {\tt MaMuPaXS} development, but at this stage we consider that this is beyond the scope of this work.

We will consider a process of the form $\varphi_1(k_1)\, \varphi_2(k_2) \to \phi_1(p_1)\, \phi_2(p_2) ...\, \phi_n(p_n)$. The code implements a change of variables for the 4-dimensional delta function in the phase-space integration~(\ref{eq:phasespace}),
\begin{equation}
    \delta^{(4)}\left(q-\sum_{i=1}^n p_i\right),
\end{equation}
which can be analytically integrated. The initial state will have total four-momentum $q = k_1 + k_2  = (\sqrt{s},0,0,0)$ in the center-of-mass rest frame. 

In spherical coordinates, each massless final state particle 4-momenta have the form:
\begin{equation}
    p_i= E_i 
(1,\sin\theta_i\cos\phi_i,\sin\theta_i\sin\phi_i,\cos\theta_i)\, , 
\qquad E_i\geq 0\,, 
\qquad\theta_i\in [0,\pi)\, ,  
\quad\phi_i\in [0,2\pi)\,.
\end{equation}
The argument of the 4-dimensional delta function is equal to the null four-vector if
\begin{equation}
    \left\{
    \begin{aligned}
        \sqrt{s} &= \sum_{i=1}^n E_i\\ 
        0 &= \sum_{i=1}^n E_i  
        \sin\theta_i\cos\phi_i \\
        0 &= \sum_{i=1}^n E_i 
        \sin\theta_i\sin\phi_i \\
        0 &= \sum_{i=1}^n E_i   
        \cos\theta_i
    \end{aligned}
    \right.\,,
\end{equation}
where $n$ is the number of particles in the final state and the new integration variables are the spherical angles $\{\theta_i,\phi_i\}$ and the energies $E_i$, equal to the three-momentum modulus $|\vec{p}_i|$ in the massless case studied here.

In this work, we consider two different cases. The first one, $n\geq 4$, where the previous system of equations is actually a linear system of 4 equations that can be solved analytically,
\begin{equation}
\label{eq:extractps}
    \left\{
    \begin{aligned}
        \sqrt{s} - \sum_{i=5}^n E_i &= \sum_{i=1}^4 E_i \\
        -\sum_{i=5}^n E_i\sin\theta_i\cos\phi_i&= \sum_{i=1}^4 E_i\sin\theta_i\cos\phi_i \\
        -\sum_{i=5}^n E_i\sin\theta_i\sin\phi_i &= \sum_{i=1}^4 E_i\sin\theta_i\sin\phi_i \\
        -\sum_{i=5}^n E_i\cos\theta_i &= \sum_{i=1}^4 E_i\cos\theta_i
    \end{aligned}
    \right.\,.
\end{equation}
This system has a unique solution for $E_1,\, E_2,\, E_3,\, E_4$. After using the delta-function in the phase-space integral to fix $E_1,\, E_2,\, E_3,\, E_4$, we are just left with an integration over the spherical angles $\{\theta_i,\phi_i\}$ and the remaining energies $\{E_5,E_6,\dots\}$ (if any).

Furthermore, because of the symmetry of the problem (two colliding massless scalars along the $z$-axis, producing scalar particles in the final state), the scattering is independent of the absolute azimuthal orientation: We can fix one of the $\phi_i$ angles in all the cases ($n=3,4,5,\dots$) without any loss of generality. The integration of this azimuth yields a $2\pi$ factor in the phase-space integration.

Hence, we will change the integration variables of eq.~(\ref{eq:phasespace}) to spherical coordinates,
\begin{equation}
    \prod_{j=1}^n dE_j\,d^3\vec{p_j} = \prod_{j=1}^n dE_j\,d\theta_j\,d\phi_j\,E_j^2\sin\theta_j\,.
\end{equation}
where $\theta_j\in [0,\pi)$, $\phi_j\in [0,2\pi)$ and $E_j=\lvert\vec{p}_j\rvert\in [0,\infty)$. Then, the 4-dimensional Delta must be applied. But, for $n\leq 4$, we can write
\begin{equation}\label{eq:deltaF}
    \delta^{(4)}\left( \sum_{j=1}^n\widetilde{p}_j\, -\,  q \right)  
    \, =\,  \delta^{(4)}\left(  
    F(p_1,p_2,p_3,p_4)  
    \, +\,  \sum_{j=5  
    }^n  p_j\, -\, q    
    \right)\, ,
\end{equation}
where the total four-momentum $q=(\sqrt{s},0,0,0)^T $ and the $p_j=(E_j, \vec{p}_j)^T$ must be understood as column vectors. $F:\mathbb{R}^4\to\mathbb{R}^4$ is a linear function on $E_1,\, E_2,\, E_3,\, E_4$. It can be expressed in the form $F(p_1,p_2,p_3,p_4)=A \vec{E}$, with $\vec{E}=(E_1,E_2,E_3,E_4)^T$ and
\begin{equation}
    A = \begin{bmatrix}
        1 & 1 & 1 & 1 \\
        \sin\theta_1\cos\phi_1 & \sin\theta_2\cos\phi_2 & \sin\theta_3\cos\phi_3 & \sin\theta_4\cos\phi_4 \\
        \sin\theta_1\sin\phi_1 & \sin\theta_2\sin\phi_2 & \sin\theta_3\sin\phi_3 & \sin\theta_4\sin\phi_4 \\
        \cos\theta_1           & \cos\theta_2           & \cos\theta_3           & \cos\theta_4          
    \end{bmatrix}\,.
\end{equation} 

Hence, the 4-dimensional delta can be written as
\begin{equation}
    \delta^{(4)}\left(\sum_{i=1}^n p_i\, -\, q \right) = \delta^{(4)}\left( A\vec{E} \, +\,  \sum_{j=5 
    }^n p_j \, -\, q\right)\, .
\end{equation}
The 4-dimensional delta can be computed via solving $E_1$, $E_2$, $E_3$ and $E_4$ from equation~(\ref{eq:extractps}), and multiplying the integrand by the inverse of the Jacobian, $(\det A)^{-1}$. 

\subsection{Case with \texorpdfstring{$n=3$}{n=3}}
If $n=3$, the system of 4 linear equations may be, in principle, overdetermined. Hence, one of the angles will be a function of the other angles. Actually, this additional restriction is the well-known fact that the 3-momenta of the 3 particles in the final state must be coplanar in the center-of-mass rest frame. This coplanarity compatibility requirement can be written as:
\begin{equation}
    \det %
    \begin{bmatrix}
        \sin\theta_1\cos\phi_1 & \sin\theta_2\cos\phi_2 & \sin\theta_3\cos\phi_3 \\
        \sin\theta_1\sin\phi_1 & \sin\theta_2\sin\phi_2 & \sin\theta_3\sin\phi_3 \\
        \cos\theta_1           & \cos\theta_2           & \cos\theta_3           
    \end{bmatrix} = 0\,.
\end{equation}
This equation implies an additional restriction that fixes one of the angular variables as a function of the others. The energies $E_1,\, E_2,\, E_3$ are fixed in the usual way by means of~(\ref{eq:extractps}).

In the particular case $n=3$, the $F:\mathbb{R}^4\to\mathbb{R}^4$ function that enters on eq.~(\ref{eq:deltaF}) is defined on $E_1$, $E_2$, $E_3$ and on the angle $\theta_3$. The roots of $F(E_1,E_2,E_3,\theta_3) - (\sqrt{s},0,0,0)^T$ can still be analytically found. The Jacobian of $F$ can also be computed, so that we still integrate on spherical coordinates. The only difference is that one of the spherical angles will be fixed by the Dirac delta.

\subsection{Code repository}

All the Jacobians for $n=3$, $4$ and $5$, as well as the required change of variables and the computations of $E_1$, $E_2$, $E_3$ and $E_4$ (for $n\geq 4$) or $\theta_3$ (for $n=3$) is coded in Fortran on~\cite{MaMuPaXS}. This Fortran code is called from Python via F2PY\footnote{\url{https://numpy.org/doc/stable/f2py/}}, and then integrated via VEGAS~\cite{Lepage:1977sw, Lepage:2020tgj}.

For the current version, the amplitudes must be hard-coded inside certain FORTRAN files specified in the repository documentation. Then, the FORTRAN code has to be compiled to a Python module. Afterwards, the VEGAS library for Python can be used for the numerical integration. However, since we think our code may be useful for the high energy physics community, we are planning in a future version to build a standalone FORTRAN library so that it can be used without recompiling, either within Python or linking to a Fortran program. The details about the usage of the code are explained on the GitHub repository.

\section{\texorpdfstring{SMEFT starting at $D=8$}{SMEFT starting at D=8}}
\label{app:LO-SMEFT8}

In some particular scenarios we may find that the leading corrections ($D=6$) to the flare-function are zero and the first non-vanishing corrections related to the flare function stem from $D=8$ SMEFT operators~\cite{Dawson:2022cmu}. 
For theories that have such a SMEFT description starting at $\mO(1/\Lambda^4)$, one finds, up to higher order corrections:
\begin{eqnarray}
a_1/2\, &=& \, a \, = \, 1  \, +\, \frac{d^2}{2}  \rho      \,+\,\mathcal{O}\left(d^3\right)\, ,  
\nonumber\\
a_2 \, &=& \, b \, = \, 1  \, +\, 3 d^2  \rho  \,+\,\mathcal{O}\left(d^3\right)\, ,  
\nonumber\\
a_3 \, &=&   4 d^2 \rho \,+\,\mathcal{O}\left(d^3\right)\, ,  
\nonumber\\
a_4 \, &=&    3 d^2 \rho  \,+\,\mathcal{O}\left(d^3\right)\, ,  
\nonumber\\
a_5 \, &=&   \frac{6}{5} d^2  \rho  \,+\,\mathcal{O}\left(d^3\right)\, ,  
\nonumber\\
a_6 \, &=& \frac{1}{5}d^2  \rho  \,+\,\mathcal{O}\left(d^3\right)\, , 
\label{eq:SMEFT-FF-D8}
\end{eqnarray}
where higher coefficients $a_n$ with $n\geq 7$ vanish at this order in SMEFT. 
For $\omega\omega$ scattering into two, three and four Higgs bosons, the relevant combinations of coefficients are,
\begin{eqnarray}
\hat{a}_2 &=&   2 d^2  \rho \, +\, \mO(d^3)\, ,\nonumber\\
\hat{a}_3 &=& \frac{4}{3} d^2  \rho \, +\, \mO(d^3)\, ,\nonumber\\
\hat{a}_4 &=& \frac{1}{3} d^2  \rho \, +\, \mO(d^3)\, .   
\label{eq:SMEFT-ajhat-D8}
\end{eqnarray}

In this type of scenarios, the coefficients of the flare-function follow the constraint (up to $D=10$ corrections),
\begin{eqnarray}
5 a_6 =\frac{5}{6}a_5=\frac{1}{3}a_4 =\frac{1}{4}a_3 =\frac{1}{3}\Delta a_2 = \Delta a_1\, ,   
\end{eqnarray}
with $\Delta a_1= 2\Delta a=2 c_{H\Box}^{(8)}v^4/\Lambda^4=d^2\rho$. Note that we have set to zero the terms proportional $c_{H\Box}^{(6)}$ in the result from ref.~\cite{Gomez-Ambrosio:2022qsi,Gomez-Ambrosio:2022why}. This is, in these scenarios one sets $d=d^2=0$ while keeping $d^2\rho\neq 0$.  
This type of additional SMEFT suppression is found for these couplings, for instance, in theoretical frameworks such as the 2-Higgs Doublet Model~\cite{Dawson:2022cmu}.

\section{Experimental measurements of \texorpdfstring{$\kappa_{2V}=a_2=b$}{k2V=a2=b}}
\label{app:a2-exp}

During the past recent years there has been tremendous improvement in the experimental determination of the $WWh^2$ coupling $\kappa_{2V}=a_2=b$. The study of double Higgs production at the LHC in various decay channels and with new machine learning techniques has led to an important reduction of the uncertainty.

Some of the first measurements yielded very loose allowed 95\% Confidence Level range:
\begin{eqnarray}
b &\in & [\, -1.1\, , \,  3.2\, ]
\qquad\qquad 
\mbox{CMS -- $hh\to b\bar{b}W^+W^-$ \cite{CMS:2023qiw},}   
\\
b &\in & [\, -0.4\, , \,  2.6\, ]
\qquad\qquad 
\mbox{CMS -- $hh\to b\bar{b}\tau^+\tau^-$ \cite{CMS:2022hgz},}  
\\
b &\in & [\, -1.3\, ,  \, 3.5\, ]
\qquad\qquad 
\mbox{CMS -- $hh\to b\bar{b}\gamma\gamma$ \cite{CMS:2020tkr},}     
\\
b &\in & [\, -0.55\, , \,  2.72\, ]
\qquad\quad 
\mbox{ATLAS -- $hh\to b\bar{b}b\bar{b}$ \cite{ATLAS:2020jgy}.}    
\end{eqnarray}

One then finds an additional group of experimental results with a clear improvement in the size of the experimental errors:
\begin{eqnarray}
b &\in & [\, -0.1\, ,  \, 2.2\, ]
\qquad\qquad 
\mbox{CMS -- $hh\to b\bar{b}b\bar{b}$ \cite{CMS:2022cpr},}    
\\
b &\in & [\, 0.1\, , \, 2.0\, ]
\qquad\qquad \,\,\,\, 
\mbox{ATLAS -- $hh\to b\bar{b}b\bar{b},\, b\bar{b}\tau^+\tau^-,\, b\bar{b}\gamma\gamma$ \cite{ATLAS:2022kbf},}   
\\
b &\in & [\, -0.03\, , \, 2.11\, ]
\qquad\quad 
\mbox{ATLAS -- $hh\to b\bar{b}b\bar{b}$ \cite{ATLAS:2023qzf}.}    
\end{eqnarray}

Finally, it is worth presenting the two most stringent determinations which are expected to be further confirmed and improved by future analysis:
\begin{eqnarray}
b &\in & [\, 0.62\, ,  \, 1.41\, ]
\qquad\qquad 
\mbox{CMS -- $hh\to b\bar{b}b\bar{b}$ \cite{CMS:2022gjd},}    
\\
b &\in & [\, 0.67\, , \, 1.38\, ]
\qquad\qquad 
\mbox{CMS -- all channels \cite{CMS:2022dwd},}   
\end{eqnarray}
where, in the last entry, `all channels' refer to the production channels $hh\to b\bar{b}b\bar{b},\, b\bar{b}\tau^+\tau^-,\, b\bar{b}\gamma\gamma$, multi-lepton, $b\bar{b}ZZ$.

\bibliographystyle{JHEP}
\bibliography{references}
\end{document}